\documentclass[twocolumn]{aastex631}
\usepackage{graphicx}   
\usepackage{lipsum} 
\usepackage{multirow}

\usepackage{color}

\def\mpc{\, {\rm Mpc}}

\def\hmpc{h^{-1}{\rm Mpc}}

\def\invhmpc{\;h\;{\rm Mpc}^{-1}}

\def\hkpc{h^{-1}\, {\rm kpc}}
\def\kms{\, {\rm km}\, {\rm s}^{-1}}

\def\hmsun{{h^{-1} M_{\odot}}}
\def\msun{\, M_{\odot}}
\def\mbh{\, M_{\rm BH}}
\def\mstar{\, M_{*}}
\def\gyr{\, {\rm Gyr}}

\shorttitle{CAMELS: extended model spaces}
\shortauthors{Ni et al.}
\graphicspath{{./}{figures/}}

\begin{document}

\title{The CAMELS project: Expanding the galaxy formation model space \\ with new ASTRID and 28-parameter TNG and SIMBA suites}

\correspondingauthor{Yueying Ni}
\email{yueying.ni@cfa.harvard.edu}

\author[0000-0001-7899-7195]{Yueying Ni}
\affiliation{Harvard-Smithsonian Center for Astrophysics, 60 Garden Street, Cambridge, MA 02138, USA}
\affiliation{McWilliams Center for Cosmology, Department of Physics, Carnegie Mellon University, Pittsburgh, PA 15213, USA}

\author[0000-0002-3185-1540]{Shy Genel}
\affiliation{Center for Computational Astrophysics, Flatiron Institute, 162 5th Avenue, 
New York, NY, 10010, USA}
\affiliation{Columbia Astrophysics Laboratory, Columbia University, 550 West 120th Street, 
New York, NY 10027, USA}

\author[0000-0001-5769-4945]{Daniel Anglés-Alcázar}
\affiliation{Department of Physics, University of Connecticut, 196 Auditorium Road, U-3046, Storrs, CT, 06269, USA}
\affiliation{Center for Computational Astrophysics, Flatiron Institute, 162 5th Avenue, New York, NY, 10010, USA}

\author[0000-0002-4816-0455]{Francisco Villaescusa-Navarro}
\affiliation{Center for Computational Astrophysics, Flatiron Institute, 162 5th Avenue, 
New York, NY, 10010, USA}
\affiliation{Department of Astrophysical Sciences, Princeton University, 4 Ivy Lane, Princeton, 
NJ 08544 USA}

\author[0000-0002-4728-6881]{Yongseok Jo}
\affiliation{Center for Computational Astrophysics, Flatiron Institute, 162 5th Avenue, 
New York, NY, 10010, USA}

\author[0000-0001-5803-5490]{Simeon Bird}
\affiliation{Department of Physics and Astronomy, University of California, Riverside, 900 University Ave., Riverside, CA 92521, USA}

\author[0000-0002-6462-5734]{Tiziana Di Matteo}
\affiliation{McWilliams Center for Cosmology, Department of Physics, Carnegie Mellon University, Pittsburgh, PA 15213, USA}
\affiliation{NSF AI Planning Institute for Physics of the Future, Carnegie Mellon University, Pittsburgh, PA 15213, USA}

\author[0000-0003-0697-2583]{Rupert Croft}
\affiliation{McWilliams Center for Cosmology, Department of Physics, Carnegie Mellon University, Pittsburgh, PA 15213, USA}
\affiliation{NSF AI Planning Institute for Physics of the Future, Carnegie Mellon University, Pittsburgh, PA 15213, USA}

\author[0000-0001-6627-2533]{Nianyi Chen}
\affiliation{McWilliams Center for Cosmology, Department of Physics, Carnegie Mellon University, Pittsburgh, PA 15213, USA}

\author[0000-0002-4728-6881]{Natalí S. M. de Santi}
\affiliation{Center for Computational Astrophysics, Flatiron Institute, 162 5th Avenue, 
New York, NY, 10010, USA}
\affiliation{Instituto de Física, Universidade de São Paulo, R. do Matão 1371, 05508-900, 
São Paulo, Brasil}

\author[0009-0003-0953-8931]{Matthew Gebhardt}
\affiliation{Department of Physics, University of Connecticut, 196 Auditorium Road, U-3046, 
Storrs, CT, 06269, USA}

\author[0000-0002-0152-6747]{Helen Shao}
\affiliation{Department of Astrophysical Sciences, Princeton University, 4 Ivy Lane, Princeton, 
NJ 08544 USA}

\author[0000-0001-5780-637X]{Shivam Pandey}
\affiliation{Department of Physics, Columbia University, New York, NY, USA 10027}
\affiliation{Department of Physics and Astronomy, University of Pennsylvania, Philadelphia, PA 19104, USA}

\author[0000-0001-6950-1629]{Lars \ Hernquist}
\affiliation{Harvard-Smithsonian Center for Astrophysics, 60 Garden Street, Cambridge, MA 02138, USA}

\author[0000-0003-2842-9434]{Romeel Dave}
\affiliation{Institute for Astronomy, University of Edinburgh, Royal Observatory, Edinburgh EH9 3HJ, UK}

\begin{abstract}

We present CAMELS-ASTRID, the third suite of hydrodynamical simulations in the Cosmology and Astrophysics with MachinE Learning (CAMELS) project, along with new simulation sets that extend the model parameter space based on the previous frameworks of CAMELS-TNG and CAMELS-SIMBA, to provide broader training sets and testing grounds for machine-learning algorithms designed for cosmological studies.
CAMELS-ASTRID employs the galaxy formation model following the ASTRID simulation and contains 2,124 hydrodynamic simulation runs that vary 3 cosmological parameters ($\Omega_m$, $\sigma_8$, $\Omega_b$) and 4 parameters controlling stellar and AGN feedback. 
Compared to the existing TNG and SIMBA simulation suites in CAMELS, the fiducial model of ASTRID features the mildest AGN feedback and predicts the least baryonic effect on the matter power spectrum. 
The training set of ASTRID covers a broader variation in the galaxy populations and the baryonic impact on the matter power spectrum compared to its TNG and SIMBA counterparts, which can make machine-learning models trained on the ASTRID suite exhibit better extrapolation performance when tested on other hydrodynamic simulation sets. 
We also introduce extension simulation sets in CAMELS that widely explore 28 parameters in the TNG and SIMBA models, demonstrating the enormity of the overall galaxy formation model parameter space and the complex non-linear interplay between cosmology and astrophysical processes.
With the new simulation suites, we show that building robust machine-learning models favors training and testing on the largest possible diversity of galaxy formation models.
We also demonstrate that it is possible to train accurate neural networks to infer cosmological parameters using the high-dimensional TNG-SB28 simulation set.

\end{abstract}


\section{Introduction} 
\label{sec:intro}


Traditional methods used to extract information from cosmological surveys typically rely on studying the properties of the Universe (e.g. galaxy distribution, neutral hydrogen distribution) on sufficiently large scales so that uncertainties from astrophysical phenomena such as feedback from supernovae (SNe) and supermassive black holes (SMBHs) remain small and under control. 
This usually means discarding a very large amount of data that could yield tighter constraints on the values of the cosmological parameters if interpreted correctly in light of those astrophysical phenomena.

Recent studies have shown that extracting information embedded in mildly non-linear scales can significantly tighten the constraints on the cosmological parameters \citep[e.g.][]{Seljak2017JCAP...12..009S,Hahn2020JCAP...03..040H,Banerjee2021MNRAS.500.5479B}. 
Ideally, we would like to extract the maximum available information from cosmological surveys. 
In order to achieve that, however, we need to overcome two important obstacles: 1) the optimal estimator that can capture the statistical properties of the considered tracer (e.g. galaxies) is unknown, and 2) astrophysical processes alter the properties and spatial distribution of tracers in an unknown manner. 

Recent advances in machine learning (ML) have enabled new possibilities to tackle this problem. 
On one hand, neural networks can be used to build surrogate models for summary statistics such as the power spectrum, allowing predictions to be made for different cosmologies \citep{Paco_2020}. 
On the other hand, we can attempt to marginalise uncertainties from baryonic effects by training networks based on dark matter, gas and stellar fields from a variety of simulations \citep{Villaescusa-Navarro2021ApJ...915...71V, Paco_robust_maps}. 
Thus, it seems that a powerful way to approach this problem will be to train models based on a large variety of simulations that explore different galaxy formation models spanning the possible space of cosmological and astrophysical parameters.

This is one of the main ideas behind the Cosmology and Astrophysics with MachinE Learning Simulations (CAMELS) project \citep{Villaescusa-Navarro2021ApJ...915...71V}. 
CAMELS contains thousands of N-body and state-of-the-art hydrodynamic simulations that cover a range of cosmological and astrophysical parameters.
Its data has been used for a large variety of tasks from weighing the Milky Way \citep{Pablo_2021} to checking the robustness of halo finders \citep{Valles_2022}. 
The hydrodynamic simulation suites in CAMELS contain two distinct sub-grid physics models based on IllustrisTNG \citep[henceforth TNG;][]{Weinberger2017,PillepichA_18} and SIMBA \citep{Dave2019MNRAS.486.2827D}, run with the AREPO \citep{Arepo} and GIZMO \citep{Hopkins2015_Gizmo} codes, respectively.
This allows us to test the robustness of developed ML models; i.e. to investigate whether a model trained on one simulation suite (produced by a particular subgrid physics model) will still perform well when tested on the other. 
We note that it is essential to distinguish between the precision and accuracy of the ML model. 
Models that are precise but not accurate are not useful. 
CAMELS aims to build accurate ML models by providing many different simulations that can be used to test the model's robustness (or accuracy).

Some works have shown that certain ML models are not robust to variations in subgrid physics \citep[see e.g.][]{Villaescusa-Navarro2021ApJ...915...71V, Pablo_2022,Delgado2023}, while others have found the opposite \citep{Helen_2022, Jay_2022, Pablo_2022a, Paco_robust_maps, Jay_2022a}. 
Several natural questions arise given this situation. 
Will the ML models that are robust across the two subgrid physics implementations also be accurate when tested on a third different subgrid physics model? 
Can we build more robust ML models when training on simulations from two different subgrid models combined?
Should we build ML models based on extensions of the original galaxy formation models to larger parameter spaces that capture more of their innate flexibility?

Those essential questions motivate us to expand the CAMELS simulation sets by incorporating new suites of hydrodynamic simulations with distinct subgrid physics models, as well as extend the previous simulation model parameter space to higher dimensions to explore other cosmological and astrophysical parameters that can play crucial roles in various observational properties.
Ideally, ML models built to marginalize over astrophysics uncertainties should be tested on simulations from as many galaxy formation models (with well-explored parameter spaces) as possible to gain confidence that those ML models will perform well when tested on real data.

In this paper, we introduce the third hydrodynamic simulation suite in CAMELS: CAMELS-ASTRID, carried out with the galaxy formation models of the ASTRID simulation and distinct from the models of TNG and SIMBA. 
The ASTRID simulation is a recently developed large volume, high resolution cosmological hydrodynamic simulation. 
The production run of ASTRID evolves a $(250\, \hmpc)^3$ cosmic volume with $2\times 5500^3$ particles \citep{Ni-Asterix, Bird-Asterix}, currently run to $z=1.3$. 
ASTRID is descended from the simulation and model code used to run the BlueTides \citep{Feng2016} and MassiveBlack-II simulations \citep{Khandai:2015}, and incorporates a full array of modern subgrid models for stellar and black hole feedback. 
ASTRID uses the simulation code MP-Gadget, an extremely scalable thread-based variant of Gadget which incorporates the hierarchical gravitational algorithm from Gadget-4 \citep{Springel:2021} and scales to $5\times 10^5$ cores.

We also introduce the CAMELS extension simulation sets of TNG, SIMBA, and ASTRID, which aim to further broaden the study of cosmological and astrophysical parameters in their respective galaxy formation models, in addition to the 6 fiducial parameters varied in the core CAMELS simulation sets. 
This incorporates the new TNG-SB28 set that contains 1024 simulations varying 28 parameters in the TNG model, accompanied by the TNG-1P-28 set that varies one parameter at a time, and an analogous SIMBA-1P-28 set based on the SIMBA model. 
It also incorporates the ASTRID-SBOb set that contains 1024 simulations varying $\Omega_b$ in addition to the 6 fiducial parameters, focusing on disentangling the effect of $\Omega_b$ with $\Omega_m$ for cosmological studies based on baryonic observables.


This paper is organized as follows. 
In Section~\ref{sec:method}, we first summarize the astrophysical subgrid models employed in CAMELS-ASTRID, and then give an overview of the new CAMELS-ASTRID suite and the CAMELS parameter extension simulation sets. 
Section~\ref{section: CV-gas} presents an illustrative comparison between the fiducial model of the ASTRID simulation and that of the TNG and SIMBA simulations.
In Section~\ref{section: 1P-set}, we diagnose the effect of the 6 fiducial cosmological and astrophysical parameters varied in CAMELS-ASTRID and compare them to TNG and SIMBA. 
We give a detailed review of the 28 parameters that are varied in the new TNG- and SIMBA-based simulations in the Appendix.
Section~\ref{section: CVLH} provides an overview of some cosmological and astrophysical properties covered by the five large simulation sets in CAMELS: the three LH sets of ASTRID, TNG and SIMBA, and the two parameter extension sets of ASTRID-SBOb and TNG-SB28.
In Section~\ref{Sec:tests}, we present some test cases and examples of ML applications enabled by introducing the large hydrodynamic simulation suites of CAMELS-ASTRID and TNG-SB28.
We finally summarize this work in Section \ref{sec:conclusions}.


\section{Simulations} 

In this section, we give a brief overview of the new simulation sets brought by the ASTRID astrophysical models.
We also introduce additional simulation sets in CAMELS that aim to explore an extended parameter space in the TNG and SIMBA models. 
This section is organized as follows.
Section~\ref{subsection:2.1} introduces the astrophysical models and the varied parameters in CAMELS-ASTRID. 
Section~\ref{subsection:2.2} gives an overview of the main simulation sets of CAMELS-ASTRID.
Section~\ref{subsection:2.3} gives a brief introduction of the CAMELS extension simulation sets for extended cosmological and astrophysical parameter studies.

\label{sec:method}

\subsection{ASTRID models in CAMELS} 
\label{subsection:2.1}

The ASTRID galaxy formation model is described in \cite{Bird-Asterix,Ni-Asterix}.
ASTRID utilizes a new version of the \texttt{MP-Gadget} simulation code, a massively scalable version of the cosmological structure formation code Gadget-3 \citep{Springel:2005}, to solve the gravity (with an N-body tree-particle-mesh approach; TreePM), hydrodynamics (with Smoothed Particle Hydrodynamics method), and astrophysical processes with a series of subgrid models.
An earlier version of \texttt{MP-Gadget} was used to run the \texttt{BlueTides} simulation \citep{Feng2016}, and the most recent version was used for ASTRID.

Radiative cooling and photoionization heating include primordial radiative cooling \citep{Katz1996ApJS..105...19K}, metal line cooling with the gas and stellar metallicities traced following \cite{Vogelsberger:2014}, a spatially-uniform ionizing background from \cite{FG2020} and hydrogen self-shielding following \cite{Rahmati:2013}.
Star formation is implemented based on the multi-phase star formation model in \cite{SH03}, and accounts for the effects of molecular hydrogen based on $\rm H_2$ fraction calculated from the metallicity and local column density~\citep{Krumholz2011ApJ...729...36K}.
ASTRID tracks metal enrichment from AGB stars, Type II SNe, and Type Ia SNe, following
9 individual elements (H, He, C, N, O, Ne, Mg, Si, Fe).

Galactic winds driven by stellar feedback are implemented kinetically via temporarily hydrodynamically decoupled particles.
Winds are sourced by newly formed star particles, which randomly pick gas particles from within their SPH smoothing length to become wind particles. 
Once a particle is in the wind, it is hydrodynamically decoupled for the minimum of $60$ Myr or $20 \mathrm{kpc} / v_w$, and is recoupled once its density drops by a factor of $10$. 
Particles in the wind do not experience or produce pressure forces, but they do receive the mass return, cool, and contribute to density estimates. 

In the CAMELS-ASTRID suite, we use two parameters $A_{\rm SN1}$ and $A_{\rm SN2}$ to vary the strength of the SN wind feedback. 
In particular, $A_{\rm SN1}$ modulates the total energy injection rate per unit star-formation, and $A_{\rm SN2}$ controls the speed of the SN wind. 
In the fiducial ASTRID model, the prescribed wind speed $v_w$ is proportional to the local one-dimensional dark matter velocity dispersion $v_{\rm w,fid} = \kappa_{\omega} \sigma_\mathrm{DM}$ with $\kappa_{\omega} = 3.7$  \citep[following the Illustris model, see][]{Vogelsberger2013}.
In the ASTRID suite, we have
\begin{equation}
v_w = A_{\rm SN2} \times v_{\rm w,fid}.
\end{equation}
The SN feedback model in ASTRID is purely energy-driven. 
Therefore, the asymptotic mass loading factor of the SN wind scales with the wind speed by $\eta_w \propto v_w^{-2}$.
In the fiducial ASTRID model, we have $\eta_{\rm w,fid} = ( v_{\rm w} / \sigma_{\rm 0,fid}  )^{-2}$ with $\sigma_{\rm 0,fid} = 353 $ km/s \citep{Bird-Asterix}.
With $A_{\rm SN1}$ modulating the power of the SN feedback energy, the SN wind mass loading factor in the ASTRID suite is
\begin{equation}
\eta_{\rm w} \equiv \frac{\dot{M}_w}{\dot{M}_{\rm SFR}} = A_{\rm SN1} \times \eta_{\rm w,fid} = A_{\rm SN1} \times \left(\frac{\sigma_{\rm 0, fid}}{v_w}\right)^{2}
\end{equation}
We note that $A_{\rm SN1}$ in CAMELS-ASTRID controls the total energy injection rate (power) per unit star formation, similar to the $A_{\rm SN1}$ parameter applied in the CAMELS-TNG suite.

The supermassive black hole (SMBH) models in ASTRID are described in \cite{Ni-Asterix}.
They are built upon the earlier models applied in \texttt{BlueTides} \citep{Feng2016}, and have added new features related to dynamical friction to improve the black hole (BH) dynamics and mergers as well as applied a power-law seeding prescription with BH seed mass $M_{\rm sd}$ between $3\times10^{4} \hmsun$ and $3\times10^{5} \hmsun$. 
However, due to the low resolution of CAMELS, we did not adopt the BH seeding and dynamical friction model as applied in the ASTRID production run. 
Instead, we adopted the BH seeding and dynamics prescriptions applied in \texttt{BlueTides}.
BHs with initial mass $M_{\rm sd} = 5 \times 10^5 \hmsun$ are seeded in halos with $M_{\rm h} = 5 \times 10^{10} \hmsun$ by means of the on-the-fly FOF halo finding. 
BH particles are re-positioned to the location of the local potential minimum at each active time step, and two BHs located within $2 \epsilon_g$ (where $\epsilon_g$ is the gravitational softening length) of each other are instantaneously merged. 
The gas accretion rate onto the BH is estimated via the Bondi-Hoyle-Lyttleton-like prescription \citep{DSH2005}.
We allow for short periods of super-Eddington accretion in the simulation but limit the accretion rate to $2$ times the Eddington accretion rate.
The BH radiates with a bolometric luminosity $L_{\rm Bol}$ proportional to the accretion rate $\dot{M}_{\rm BH}$, with a mass-to-light conversion efficiency $\eta=0.1$ in an accretion disk according to \cite{Shakura1973}.

SMBH feedback follows a two-mode approach including the thermal and kinetic feedback mode delineated by the Eddington ratio of the instantaneous BH accretion rate.
The Eddington threshold $\chi_{thr}$ is capped at $\chi_{\text{thr,max}} = 0.05$ and is also a function of the BH mass such that the kinetic mode is turned on only for massive BHs with $\mbh \gtrsim 5 \times 10^{8} \hmsun$.
In CAMELS-ASTRID, we use $A_{\rm AGN1}$ and $A_{\rm AGN2}$ to modulate the efficiency of the kinetic and thermal feedback separately, where $A_{\rm AGN1}$ has the same physical meaning as that in the CAMELS-TNG suite. 
In the high accretion mode, the AGN feedback is deposited thermally with
\begin{equation}
    \Delta \dot{E}_{\text {high }}= A_{\rm AGN2} \times \epsilon_{\mathrm{f,th}} \epsilon_{\mathrm{r}} \dot{M}_{\mathrm{BH}} c^{2}   \,\,\,\,\, (\lambda_{Edd} > \chi_{thr})
\end{equation}
In the fiducial run, we have a mass-to-light conversion efficiency $\epsilon_{\mathrm r} = 0.1$ and $\epsilon_{\mathrm{f,th}} = 0.05$, assuming that 5\% of the radiation energy is thermally injected to the surrounding gas within a feedback sphere with radius of two times the SPH kernel. 

The formalism of the AGN kinetic feedback in the low accretion mode largely follows \citet{Weinberger2017} with different choices of parameter values.
The AGN kinetic feedback energy is deposited as
\begin{equation}
   \Delta \dot{E}_{\text {low }} = A_{\rm AGN1} \times \epsilon_{\mathrm{f, kin}} \dot{M}_{\mathrm{BH}} c^{2}  \,\,\,\,\, (\lambda_{Edd} < \chi_{thr})
\end{equation}
where $\epsilon_{\mathrm{f, kin}}$ scales with the BH local gas density and has a maximum value of $\epsilon_{\mathrm{f, kin,max}}=0.05$. 
The energy is accumulated over time and released in a bursty way once the accumulated kinetic feedback energy exceeds the threshold $E_{\mathrm{inj}, \mathrm{min}}=f_{\mathrm{re}} \frac{1}{2} \sigma_{\mathrm{DM}}^{2} m_{\mathrm{enc}}$. 
Here $\sigma_{\mathrm{DM}}^{2}$ is the one-dimensional dark matter velocity dispersion around the BH, $m_{\mathrm{enc}}$ is the gas mass in the feedback sphere, and $f_{\mathrm{re}}=5$ for the fiducial run. 
The released kinetic energy kicks each gas particle in the feedback kernel in a random direction with a prescribed momentum weighted by the SPH kernel. 
Compared to TNG \citep{Weinberger2017}, ASTRID turns on the AGN kinetic feedback based on a more stringent criterion (with a lower $\chi_{\rm thr}$ and a higher $M_{\rm BH,pivot}$), and also adopts a lower upper limit for the feedback efficiency $\epsilon_{\mathrm{f, kin, max}}$.
Therefore, we note that the AGN kinetic feedback in the ASTRID suite is milder compared to TNG, as will be discussed in the next sections.

Table~\ref{tab:table1} briefly summarizes the physical meaning of the four astrophysical feedback parameters. 
$A_{\rm SN1}$, $A_{\rm SN2}$, and $A_{\rm AGN1}$ in ASTRID modulate the energy of SN feedback per unit star formation, the wind speed, and AGN kinetic feedback energy per unit BH accretion (similarly to those parameters in the TNG suite).
$A_{\rm AGN2}$ modulates the AGN thermal feedback energy per unit BH accretion, as the thermal mode is the dominant channel of AGN feedback in the ASTRID simulation.

We note again that the CAMELS-ASTRID suite has applied some adaptations in astrophysical models compared to the production run of ASTRID, mainly due to the limited volume and resolution of the CAMELS simulations. 
In particular, simulations in CAMELS-ASTRID do not adopt the BH power-law seeding and dynamical friction model as applied in the ASTRID production run \citep{Ni-Asterix}, as those models require higher mass resolution. 
Moreover, with the limited volume of 25 $\hmpc$, CAMELS-ASTRID does not model patchy hydrogen reionization, helium reionization, as well as the effect of massive neutrinos as in the ASTRID production run.

\begin{table*}
\centering
\begin{tabular}{| l | p{29mm}| p{23mm}| p{45mm}| p{45mm}| }
    \hline
    Simulation & $A_{\rm SN1}$  & $A_{\rm SN2}$ & $A_{\rm AGN1}$ & $A_{\rm AGN2}$  \\
    \hline
    \hline
    
    ASTRID & Galactic winds: \newline energy per unit SFR \newline [0.25 - 4.00]
            & Galactic winds: \newline wind speed \newline [0.50 - 2.00]
            & Kinetic mode BH feedback: \newline energy per unit BH accretion \newline [0.25 - 4.00]
            & Thermal mode BH feedback: \newline energy per unit BH accretion \newline [0.25 - 4.00]
            \\

    \hline
    TNG & Galactic winds: \newline energy per unit SFR \newline [0.25 - 4.00]
                 & Galactic winds: \newline wind speed \newline [0.50 - 2.00]
                 & Kinetic mode BH feedback: \newline energy per unit BH accretion \newline [0.25 - 4.00]
                 & Kinetic mode BH feedback: \newline ejection speed / burstiness \newline [0.50 - 2.00]
                \\

    \hline
    SIMBA & Galactic winds: \newline mass loading \newline [0.25 - 4.00]
          & Galactic winds: \newline wind speed \newline [0.50 - 2.00]
          & QSO \& jet-mode BH feedback: \newline momentum flux \newline [0.25 - 4.00]
          & Jet-mode BH feedback: \newline jet speed \newline [0.50 - 2.00]
          \\
    
    \hline  
\end{tabular}
\caption{This table summarizes the physical meaning of the four astrophysical parameters ($A_{\rm SN1}$, $A_{\rm SN2}$, $A_{\rm AGN1}$, $A_{\rm AGN2}$) in the ASTRID, TNG, and SIMBA suites. The fiducial parameter value in each simulation is normalized to $A_{\rm SN1}$ = $A_{\rm SN2}$ = $A_{\rm AGN1}$ = $A_{\rm AGN2}$ = 1. The variation of each parameter is also shown in each cell. We note that the range of $A_{\rm AGN2}$ in the ASTRID suites is different from TNG and SIMBA, as $A_{\rm AGN2}$ in the ASTRID suite represents energy flux, similar to $A_{\rm AGN1}$. }
\label{tab:table1}
\end{table*}

\begin{table*}
\centering
\begin{tabular}{| c | c | c | c | c | c | }
    \hline
    Name & Hydrodynamic Method & Code & Simulation Set & Number & Varying parameters  \\
    \hline
    \hline
    \multirow{5}{*}{ASTRID} 
    & \multirow{5}{*}{Pressure-Entropy SPH}
    & \multirow{5}{*}{MP-Gadget} 
    & CV & 27 & S \\
    & & & 1P & 61 & $\Omega_m$, $\sigma_8$, $A_{\rm SN1}$, $A_{\rm SN2}$, $A_{\rm AGN1}$, $A_{\rm AGN2}$ \\
    & & & LH & 1000 & $\Omega_m$, $\sigma_8$, $A_{\rm SN1}$, $A_{\rm SN2}$, $A_{\rm AGN1}$, $A_{\rm AGN2}$, S \\
    & & & EX  & 4 & $A_{\rm SN1}$, $A_{\rm SN2}$, $A_{\rm AGN1}$, $A_{\rm AGN2}$ \\
    \hline
\end{tabular}
\caption{Summary of the core ASTRID simulation sets, which share the same design as CAMELS-TNG and CAMELS-SIMBA, as described in \cite{Villaescusa-Navarro2021ApJ...915...71V} (see their Table 2). $A_{\rm SN1}$, $A_{\rm SN2}$, $A_{\rm AGN1}$, $A_{\rm AGN2}$ represent the value of subgrid physics parameters controlling stellar and AGN feedback. $S$ is the initial random seed of the initial conditions of a simulation. 
The LH set is a Latin hypercube where the values of \{$\Omega_m$, $\sigma_8$, $A_{\rm SN1}$, $A_{\rm SN2}$, $A_{\rm AGN1}$, $A_{\rm AGN2}$, S\} are varied simultaneously. 
Note that the parameters in the Latin hypercube are different between ASTRID, TNG, and SIMBA. 
Simulations in the 1P set have the same initial random seed and vary only one of the parameters at a time. Simulations in the CV set have fixed values of the cosmological and astrophysical parameters (at fiducial values) but different initial random seeds. The EX set has the simulations with the same cosmological parameter and random seeds while exploring the extreme values in the 4 astrophysical parameters. See text for further details.}
\label{tab:table2}
\end{table*}

\begin{table*}
\centering
\begin{tabular}{| c | c | c | c | }
    \hline
    Name & Simulation Set & Number & Varying parameters  \\
    \hline
    \multirow{2}{*}{ASTRID Extension}  
    & ASTRID-SBOb & 1024 & \{$\Omega_m$, $\sigma_8$, $A_{\rm SN1}$, $A_{\rm SN2}$, $A_{\rm AGN1}$, $A_{\rm AGN2}$\} + $\Omega_b$, S \\
    & ASTRID-1P-Ob & 8 & $\Omega_b$ \\
    \hline
    \multirow{2}{*}{TNG Extension}  
    & TNG-SB28 & 1024 & \{$\Omega_m$, $\sigma_8$, $A_{\rm SN1}$, $A_{\rm SN2}$, $A_{\rm AGN1}$, $A_{\rm AGN2}$\} + 22 additional parameters, S \\
    & TNG-1P-28 & 88 & 22 additional parameters  (details in Appendix~\ref{sec:appendix-28param}) \\
    \hline
    {SIMBA Extension}  
    & SIMBA-1P-28 & 88 & 22 additional parameters  (details in Appendix~\ref{sec:appendix-28param}) \\
    \hline
\end{tabular}
\caption{
Summary of the additional simulation sets in CAMELS (based on ASTRID, TNG and SIMBA) for extended parameter studies. 
The ASTRID-SBOb set is the extension of the ASTRID-LH set that also has $\Omega_b$ varied in addition to the original varied 6 parameters, where the 7 parameters are sampled in a Sobol sequence.
TNG-SB28 set is the extension of the TNG-LH set that explores the variation of 22 additional cosmological and astrophysical parameters, with individual parameter variation studied in the TNG-1P-28 set. 
The SIMBA-1P-28 simulation set presents individual parameter variations in the SIMBA model for 22 additional cosmological and astrophysical parameters.
See text for further details.}
\label{tab:table3}
\end{table*}

\subsection{CAMELS-ASTRID Suite}
\label{subsection:2.2}

CAMELS-ASTRID is a third hydrodynamic simulation suite in CAMELS \citep{Villaescusa-Navarro2021ApJ...915...71V} brought by the astrophysical models of the ASTRID simulation. 
The core CAMELS-ASTRID suite shares a similar design to the simulation sets from the 
``TNG" and ``SIMBA" suites, containing 1092 simulation runs in 4 simulation sets (CV, 1P, LH, EX) as will be briefly discussed below.
In Table~\ref{tab:table2}, we summarize the core simulation sets of the CAMELS-ASTRID suite (appending to Table 2 of \citealp{Villaescusa-Navarro2021ApJ...915...71V}) and describe them individually below.
We introduce the extended simulation sets of CAMELS (based on ASTRID, TNG and SIMBA) in the next subsection.

(1) The CV set (CV for cosmic variance) contains 27 simulations with different realizations (random seed) of initial conditions and fiducial cosmological and astrophysical parameters fixed at $\Omega_m = 0.3$, $\sigma_8 = 0.8$, $A_{\rm SN1} = A_{\rm SN2} = A_{\rm AGN1} = A_{\rm AGN2} = 1$. 
Note that the CV set shares the same set of the initial random seeds across ASTRID, TNG, and SIMBA, allowing cross-comparison between the three simulation suites of the same realizations with their respective fiducial models. 

(2) The 1P set (1P for 1-parameter) contains 61 simulations sharing the same realization (random seed) of the initial conditions and varying the cosmological and astrophysical parameters one at a time.
The varied ranges of the parameters are $\Omega_m \in [0.1, 0.5]$, $\sigma_8 \in [0.6, 1.0]$, $A_{\rm SN1} \in [0.25, 4.0]$, $A_{\rm SN2} \in [0.5, 2.0]$, $A_{\rm AGN1} \in [0.25, 4.0]$, and $A_{\rm AGN2} \in [0.25, 4.0]$.
The spacing is linear for $\Omega_m$ and $\sigma_8$ and logarithmic for the 4 astrophysical parameters. 

We note that the different range of $A_{\rm AGN2}$, which is varied between [0.25 - 4.00] for ASTRID, and between [0.5 - 2.0] for TNG and SIMBA, is motivated by their corresponding physical meaning. 
Unlike $A_{\rm AGN2}$ in the TNG and SIMBA suites that modulates the gas ejection speed in the jet/kinetic mode feedback, the function of $A_{\rm AGN2}$ in ASTRID resembles that of $A_{\rm AGN1}$ and controls the AGN feedback efficiency in another channel (thermal feedback), therefore it shares the same variation range as $A_{\rm AGN1}$.

(3) The LH set (LH for Latin hypercube) contains 1000 simulations with the value of the cosmological and astrophysical parameters ($\Omega_m$, $\sigma_8$, $A_{\rm SN1}$, $A_{\rm SN2}$, $A_{\rm AGN1}$, $A_{\rm AGN2}$) arranged in a Latin hypercube with the same parameter range described for the 1P set. The initial random seed is also different for each simulation.

(4) The EX set (EX for extreme) contains 4 simulations with the same initial random seed and fiducial cosmological parameters $\Omega_m = 0.3$, $\sigma_8 = 0.8$. One simulation has the fiducial astrophysical parameters;
one has extreme SN feedback with $A_{\rm SN1} = 100$;
one has extreme kinetic AGN feedback with $A_{\rm AGN1} = 100$;
and the last one has no SN or AGN kinetic feedback with $A_{\rm SN1} = A_{\rm AGN1} = 0$.

We note again that the CV, 1P, and EX sets of ASTRID share the same realization of the initial conditions as their counterparts used in the TNG and SIMBA suites (for example, the three simulation suites share the same initial conditions for CV0), while the LH and SBOb sets (see Section \ref{subsection:2.3}) use a separate set of random seeds to generate the initial conditions.

As in the CAMELS-TNG and CAMELS-SIMBA suites, each simulation in the CAMELS-ASTRID suite evolves a periodic box of comoving volume equal to ($25 \hmpc$)$^3$ from $z=127$ to $z=0$, with $256^3$ dark matter particles and $256^3$ gas particles in the initial conditions. 
In the core simulation sets described above, each simulation shares the cosmological parameters of $h=0.6711$, $n_s=0.9624$, $M_{\nu} = 0.0$ eV, $\omega=-1$, $\Omega_K = 0$ and $\Omega_{b} = 0.049$.

The simulations in the CAMELS-ASTRID suite each contain 91 snapshots from $z=15$ to $z=0$, covering the time stamps in the TNG and SIMBA suites but with higher time resolution for merger tree generation. 
For each snapshot, dark matter halos and subhalos/galaxies are identified using the SUBFIND \citep{Springel2001MNRAS.328..726S}, Rockstar \citep{Behroozi2013ApJ...762..109B} and AMIGA \citep{Knollmann2009ApJS..182..608K} halo finders.
Moreover, we also have merger trees generated from Rockstar \citep{Behroozi2013_MergerTree} and SubLink \citep{Srisawat2013MNRAS.436..150S} available for each simulation.

\begin{figure*}
\centering
\includegraphics[width=1.0\textwidth]{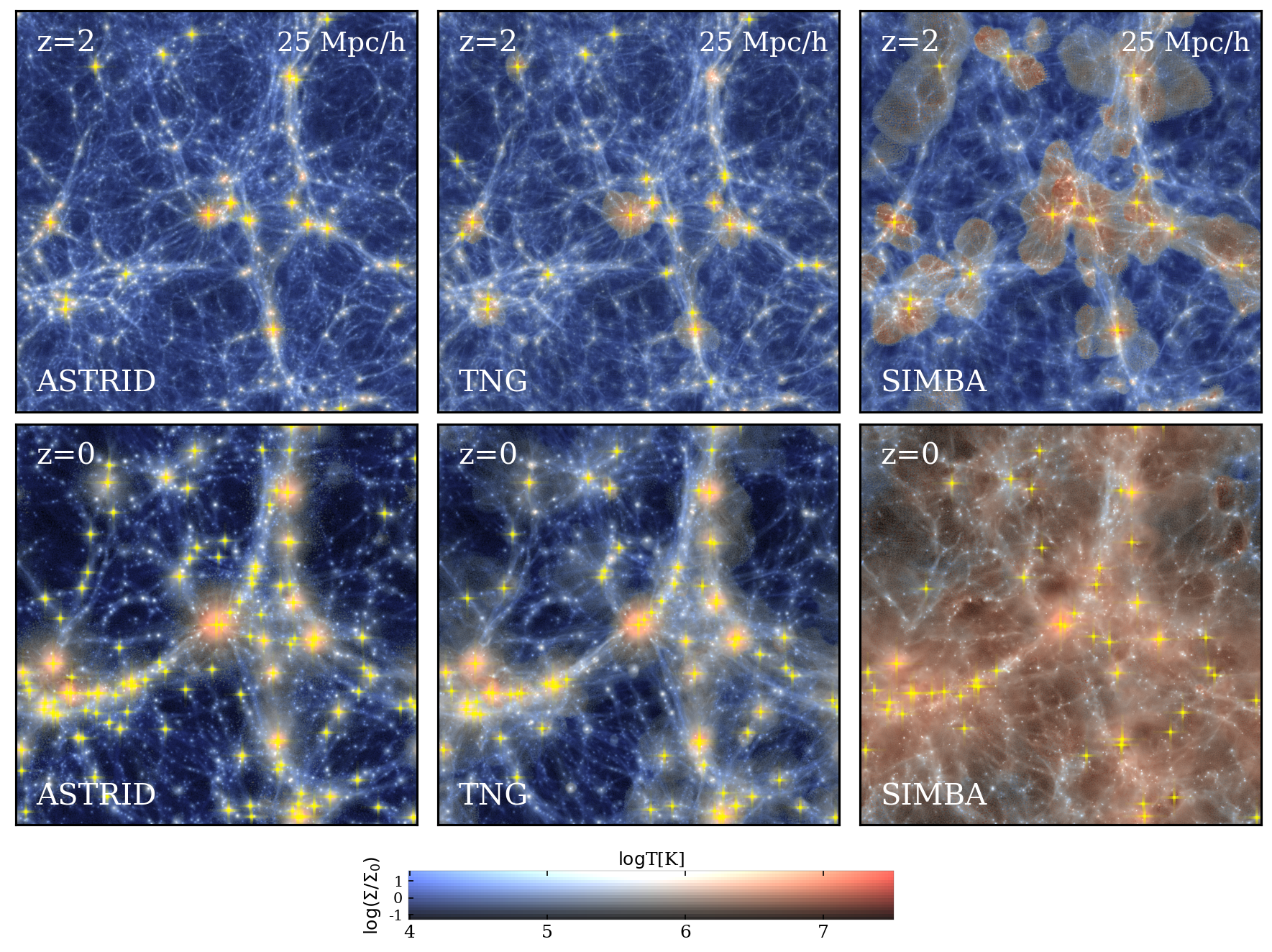}
\caption{Illustration of the gas field of CV0 set for ASTRID (the first column), TNG (the second column), and SIMBA (the third column), centered on the most massive halo in the simulation. Each panel shows the gas density field colored by temperature (blue to red indicating cold to hot respectively, as indicated by the 2D color map) over the full box volume of $(25 \hmpc)^3$ at $z=2$ (upper panel) and $z=0$ (lower panel). The yellow spikes mark the locations of massive BHs with $M_{\rm BH}>10^{8} \msun$.}
\label{fig:GasImage}
\end{figure*}

\subsection{CAMELS Extension Simulation Sets} 
\label{subsection:2.3}

The core CAMELS hydrodynamic simulation sets have 6 cosmological and astrophysical parameters varied, covering the key parameter space of interest with 1000 samples in the LH sets of ASTRID, TNG and SIMBA suites. 
However, there are still many unexplored parameters in cosmology and astrophysical models that can have a significant impact on many observational properties and can have non-negligible consequences for cosmological studies.
Sampling the higher dimensional parameter space with hydrodynamic simulations is rather expensive. 
Here we introduce the recently developed extension simulation sets in CAMELS that serve as the first stepping stone for extended parameter studies. 
Table~\ref{tab:table3} gives a summary of the extension simulation sets based on the ASTRID, TNG and SIMBA models.

\textbf{TNG extension}:
We introduce here the TNG-SB28 set, which is an analog of the original TNG-LH set except where not 6 but 28 parameters of the model are varied simultaneously. They represent a semi-complete account of the free parameters in the TNG model, and they include 5 cosmological parameters, 2 parameters concerning star-formation and the ISM, 2 parameters related to stellar population modeling, 10 parameters controlling galactic wind feedback, 3 parameters governing the growth of SMBHs, and 6 parameters describing AGN feedback. We provide concise yet well-defined descriptions of each of the 28 parameters, including their range of variation, in Appendix \ref{sec:appendix-2-tng}.

Another, minor difference between TNG-SB28 and the original TNG-LH is that the 28-dimensional parameter space is sampled using a Sobol sequence rather than a Latin hypercube. Both methods provide a quasi-uniform (or low-discrepancy), yet irregular, sampling of the space, but compared to the Latin hypercube, the Sobol sequence is deterministic, faster to compute, has a lower discrepancy, and more easily allows the potential addition of newer, future samples.

In addition to the semi-uniform sampling of the 28-dimensional parameter space, we introduce another new simulation set, TNG-1P-28, which is composed of simulations that vary one parameter at a time, iterating over each of the 22 new parameters beyond the original 6 of the TNG-LH set. In TNG-1P-28, there are four simulations around the fiducial one for each parameter, two where the value of the parameter is lower than the fiducial value and two where it is higher\footnote{There is one exception to this rule, when the fiducial value is zero and negative values are unphysical, hence the four variations all use larger values than the fiducial; see details in \ref{sec:appendix-2-tng}.}. The minimum and maximum values of each parameter are the same between TNG-1P-28 and TNG-SB28. The spacing of values for each parameter in TNG-1P-28 is uniform either in linear or in logarithmic space, and that distinction translates directly to whether the sampling of values in the Sobol sequence of TNG-SB28 is performed in linear or logarithmic space.

The range of variation of each parameter was chosen based on a combination of (sometimes competing) two considerations: {\it i)} a physical intuition into its realistic range of values, with an attempt to err on the side of a large range in order to avoid edge effects, and  {\it ii)} an empirical examination of its effects on the simulation results, with a (very) rough goal of having the variations of different parameters resulting in effects of similar magnitudes. Appendix \ref{sec:appendix-28param} presents the effects of the various parameters on a select number of physical quantities from the simulation results based on the TNG-1P-28 set. It is important to note that for any given physical quantity, some model parameters are significantly more influencial than others. We have verified, however, that each of the 28 parameters has a significant impact on at least {\it some} physical quantities that can be considered as basic to any cosmological galaxy formation model.

\textbf{ASTRID extension}:
Apart from the fiducial hydrodynamic simulation sets, the ASTRID suite also contains an additional SBOb set that adds one more degree of freedom by also varying $\Omega_{b}$ in addition to the 6 parameters in the ASTRID-LH set. 
ASTRID-SBOb contains 1024 simulations with the value of the cosmological and astrophysical parameters ($\Omega_m$, $\sigma_8$, $\Omega_b$, $A_{\rm SN1}$, $A_{\rm SN2}$, $A_{\rm AGN1}$, $A_{\rm AGN2}$) arranged in a Sobol sequence with the same parameter ranges as in the 1P set. 
The variation range of $\Omega_{b}$ is $\Omega_{b} \in [0.01, 0.09]$.
Unlike the TNG-SB28 set that broadly (and sparsely) samples 28 parameters in the simulation model, the main focus of the ASTRID-SBOb set is to explicitly disentangle the effect of $\Omega_b$ and $\Omega_m$ for cosmological parameter inference based on baryonic observables \citep[e.g.,][]{Villaescusa-Navarro2022_OneGalaxy,deSanti2023,Shao2023_symbolic}.
The initial random seed is also unique for each simulation.

\textbf{SIMBA extension}:
In analogy with the TNG-1P-28 set, we introduce the SIMBA-1P-28 simulation set to investigate the impact of 22 new cosmological and astrophysical parameters beyond the 6-parameter variations in the original CAMELS-SIMBA suite. 
The SIMBA-1P-28 set consists of 88 simulations that vary one parameter at a time, including four simulations around the fiducial model for each of the 22 new parameters (two simulations decreasing the parameter value and two simulations increasing the parameter value relative to the fiducial value).  
In full, the SIMBA-1P-28 suite includes variations of 5 cosmological parameters, 3 parameters controlling star formation and the ISM, 8 parameters controlling galactic winds driven by stellar feedback, 6 parameters controlling the growth of SMBHs, and 6 parameters controlling AGN feedback. 
We describe the parameters in detail and their range of variation in Appendix~\ref{sec:appendix-2-simba}.  We illustrate the effects of the various parameter variations on different physical quantities in Appendix~\ref{sec:appendix-SIMBA-1P}.

\section{Comparison of the three fiducial models}
\label{section: CV-gas}

\begin{figure*}
\centering
\includegraphics[width=1.0\textwidth]{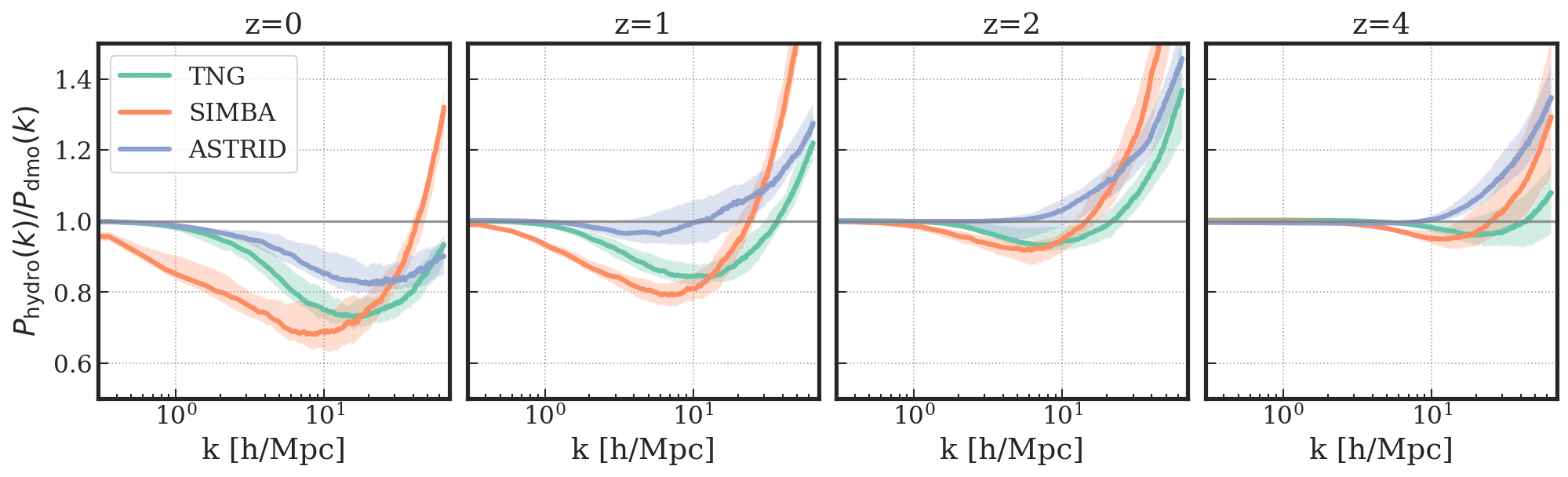}
\caption{Illustration of baryonic feedback on the total matter density field in the fiducial model of the three simulation suites. The $x$ axis is the $k$ mode in comoving units and the $y$ axis in each panel gives the ratio of the total matter density power spectrum in hydrodynamic simulations relative to that of the corresponding N-body simulations for the CV set, at redshifts $z=0,1,2,4$. 
Green, orange and blue lines represent the results from TNG, SIMBA, and ASTRID separately. The solid line and shaded regions give the median and 10-90 percentiles from the 27 simulations in the CV set.}
\label{fig:PsRatio-CV}
\end{figure*}

\begin{figure}
\centering
\includegraphics[width=0.9\columnwidth]{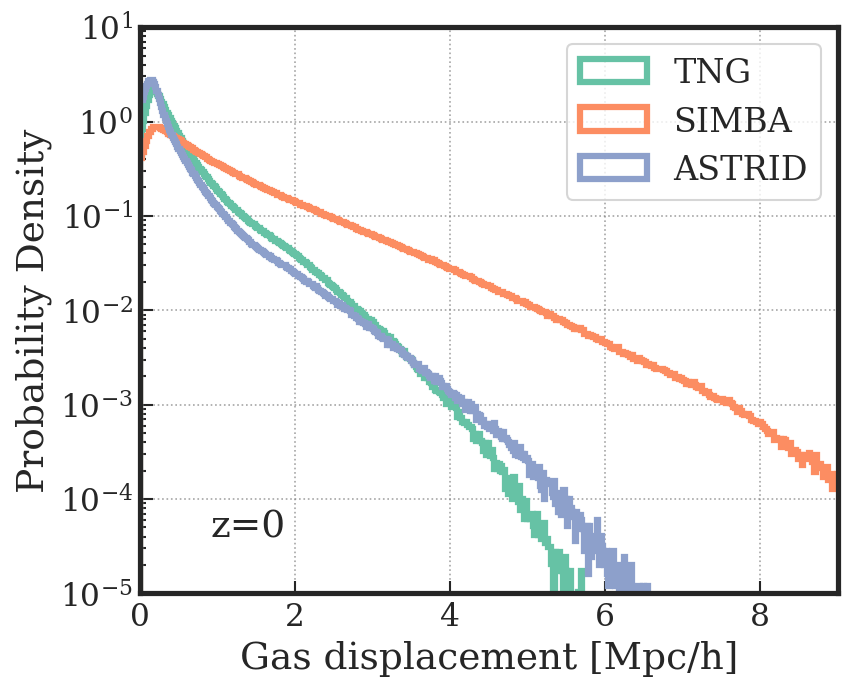}
\caption{Comparison of the spatial distribution of baryons in the fiducial model of the three simulation suites.
The $x$ axis represents the displacement of gas elements at $z=0$ relative to the initial neighboring dark matter particles from the initial conditions. The $y$ axis gives the probability distribution of the gas elements within the given displacement value.
}
\label{fig:BaryonSpread}
\end{figure}


We start with a comparison between the fiducial models of the ASTRID, TNG, and SIMBA suites, to demonstrate the systematic differences between the three hydrodynamic simulations brought by their respective sub-grid physical models. 
We highlight one of the most distinctive and interesting differences among the three simulation models: large-scale gas properties and correspondingly the influence of baryonic feedback on the matter power spectrum.

Fig.~\ref{fig:GasImage} shows the large-scale gas density field colored by the temperature at $z=2$ and $z=0$ from the CV0 simulation, where the three simulation suites share the same initial conditions and adopt their corresponding fiducial astrophysical models \footnote{see \url{https://www.youtube.com/watch?v=zfxBa_Zp6WM} for a movie visualization of the systematic difference between TNG, SIMBA, ASTRID and Magneticum in the large scale gas field.}.
One can see the clear differences with respect to the gas thermal properties of the intergalactic medium (IGM), where SIMBA heats up the gas surrounding large halos starting from higher redshift ($z \gtrsim 2$), producing hotter IGM gas compared to ASTRID and TNG. 
The striking difference in the large-scale gas field is mostly caused by the AGN feedback models, as in SIMBA AGN feedback turns on earlier and deposits kinetic and thermal energy on larger scales with fast jet outflows \citep[see][]{Dave2019MNRAS.486.2827D,Christiansen2019,Borrow2020}.  

To illustrate where the most efficient AGN feedback takes place, Fig.~\ref{fig:GasImage} marks the position of massive black holes with $\mbh > 10^{8} \msun$ in yellow spikes. 
The bipolar lobe of hot gas in SIMBA at $z=2$ clearly indicates the imprint of AGN jet mode feedback. 
On the other hand, TNG and ASTRID adopt a more localized (and also non-collimated) AGN kinetic feedback model for massive black holes at low Eddington ratio and therefore exhibit less impact on the gas field on large scales. 
Compared to TNG, ASTRID AGN kinetic feedback is even milder and turns on later (as it hinges on larger black hole masses $M_{\rm BH} > 5 \times 10^{8} \hmsun$) and therefore it has the least impact on the large scale gas properties among the three simulations. 
The gas properties have been studied previously in the work of \cite{Contreras2023MNRAS.519.2251B} for warm-hot IGM gas and \cite{Tillman2022arXiv221002467T} for the Lyman alpha forest using the TNG and SIMBA suites, and will be followed up with new simulation suites of ASTRID. 

We compare the fiducial models of ASTRID, TNG, and SIMBA and diagnose their baryonic feedback impact across different redshifts. 
In Fig.~\ref{fig:PsRatio-CV}, we quantify the baryonic feedback of the fiducial models by calculating the ratio of the total matter power spectrum from the hydrodynamic simulations to that from their respective dark matter only simulations in the CV set, where the lines show the median of $P_{\rm hydro}/P_{\rm dmo}$ for the CV set and the shaded regions give the 10-90 percentile in each $k$ bin.

The $P_{\rm hydro}/P_{\rm dmo}$ ratio clearly shows the systematic differences in baryonic feedback between the three simulation suites with their fiducial astrophysical models. 
Compared to the collisionless dark matter only simulation, gas cooling and star formation enhance matter clustering on small scales, while the astrophysical feedback processes redistribute matter (especially the baryonic gas) and suppresses clustering (as in \citealt{Springel2018MNRAS.475..676S,vanDaalen2020,Delgado2023}).
These two competing processes leave imprints on different scales and together shape the suppression and enhancement features in the $P_{\rm hydro}/P_{\rm dmo}$ ratio as seen in Fig.~\ref{fig:PsRatio-CV}.

Among different astrophysical feedback processes, the AGN jet mode is the most important channel for efficiently re-distributing the gas on large scales.
Therefore, among the three simulation suites, SIMBA with its most aggressive AGN feedback (that launches high-speed gas outflows and deposits energy at large distances that can reach hundreds of kpc away from the AGN host halos) produces the most prominent suppression of the matter power on large scales. 
On the other hand, ASTRID exhibits the least impact on the matter power spectrum as a consequence of having the mildest AGN feedback among the three simulation suites.
The power spectrum ratio at $z=0$ can reach $<70\%$ at $k \sim 10 \hmpc$ in the SIMBA model while only $90\%$ in ASTRID at the same $k$ scale.

The baryonic impact on the matter power spectrum in the three simulation models exhibits a similar trend in time evolution.
At higher redshifts ($z > 2$ in this small volume) when the majority of black holes have not grown massive enough to launch jets, baryons mostly enhance the matter power as the result of gas cooling and galaxy formation.
When going to lower redshifts, the suppression of power emerges and propagates to large scales with time, as black holes become more massive and produce more powerful feedback that can affect the matter distribution on larger scales. 
The onset of matter power suppression is earliest in SIMBA and latest in ASTRID, mainly due to the different AGN feedback models.

As an alternative metric of baryonic feedback, Fig.~\ref{fig:BaryonSpread} compares the spatial distribution of baryons at $z=0$ relative to the dark matter component in the fiducial model of the three simulation suites.  
The spread of baryons relative to dark matter is quantified in a Lagrangian sense, by calculating the distance of gas elements at $z=0$ from their original dark matter particle neighbors at the initial conditions \citep{Borrow2020,Matt-prep}. 
Overall, SIMBA displaces gas farther than TNG and ASTRID.
At $z=0$, the SIMBA model pushes 40\% of gas elements more than 1\,Mpc away from their original neighboring dark matter, while in ASTRID only $\sim$10\% of gas spreads more than 1\,Mpc away.
Compared to TNG, ASTRID on average displaces fewer baryons out to intermediate distances but can have a small number of gas particles reach more extreme distances.
A more detailed study of the baryon spread and its correspondence to the matter power spectrum is presented in \cite{Matt-prep}.

\section{1P sets: The effects of individual parameters}
\label{section: 1P-set}

\begin{figure*}
\centering
\includegraphics[width=1.0\columnwidth]{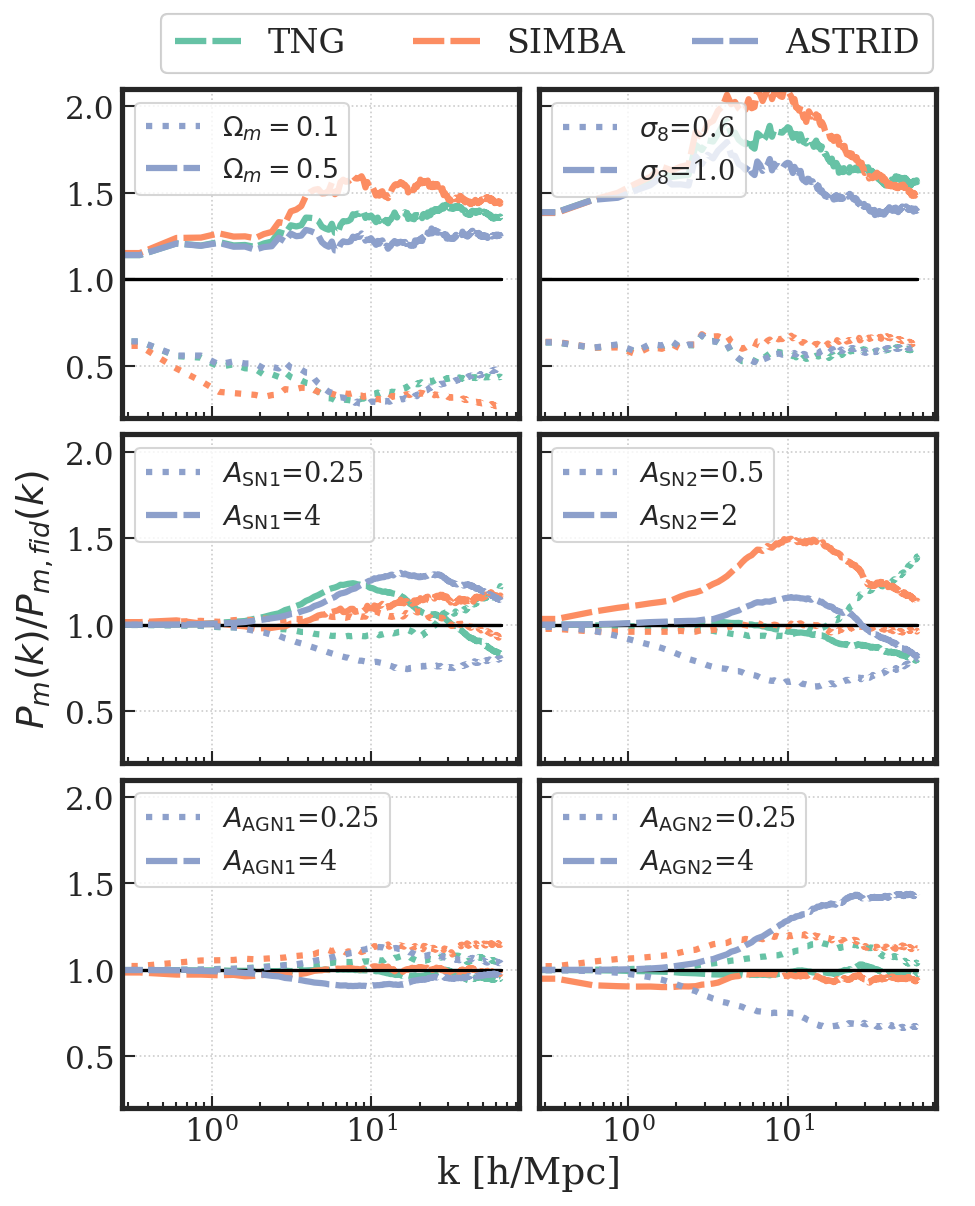}
~
\includegraphics[width=1.03\columnwidth]{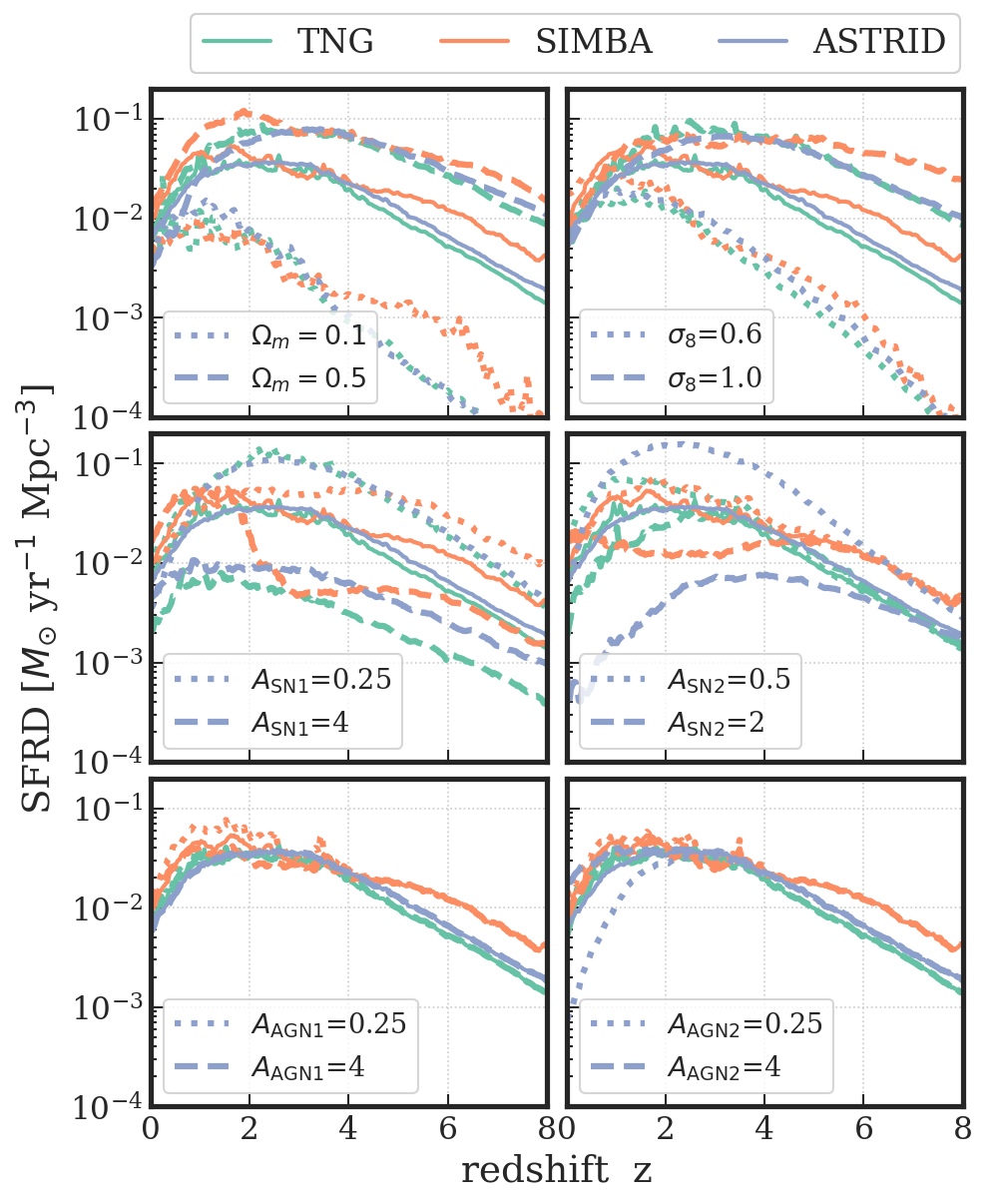}

\caption{
\textbf{Left panels:} Ratio of the matter power spectrum of the 1P simulations (where only one parameter is varied and other parameters are fixed) to that of the fiducial model. The matter power spectra are evaluated at $z=0$. Green, orange and blue colors represent the results from TNG, SIMBA, and ASTRID, respectively. The solid and dotted lines indicate the simulations with the highest and lowest parameter value. Note that the value of $A_{\rm AGN2}$ is varied between [0.25 - 4.00] for ASTRID, and between [0.5 - 2.0] for TNG and SIMBA.
\textbf{Right panels:} Global star formation rate density (SFRD) of the 1P simulations. Solid, dotted, and dashed lines represent the simulations with the fiducial, lowest, and highest parameter values in the variation range correspondingly.
}
\label{fig:1P-ps-sfrd}
\end{figure*}

\begin{figure*}
\centering
\includegraphics[width=1.0\columnwidth]{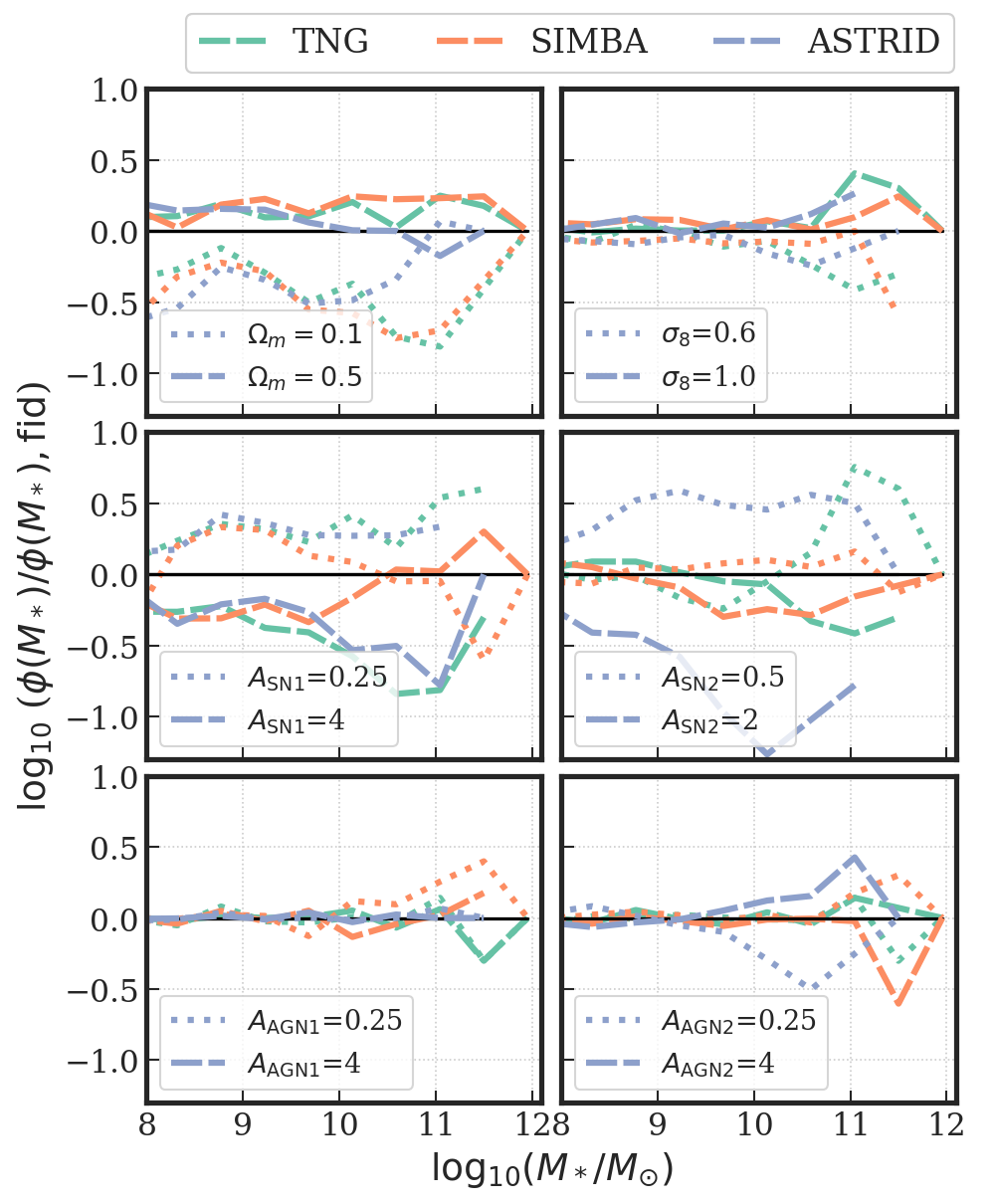}
~
\includegraphics[width=1.0\columnwidth]{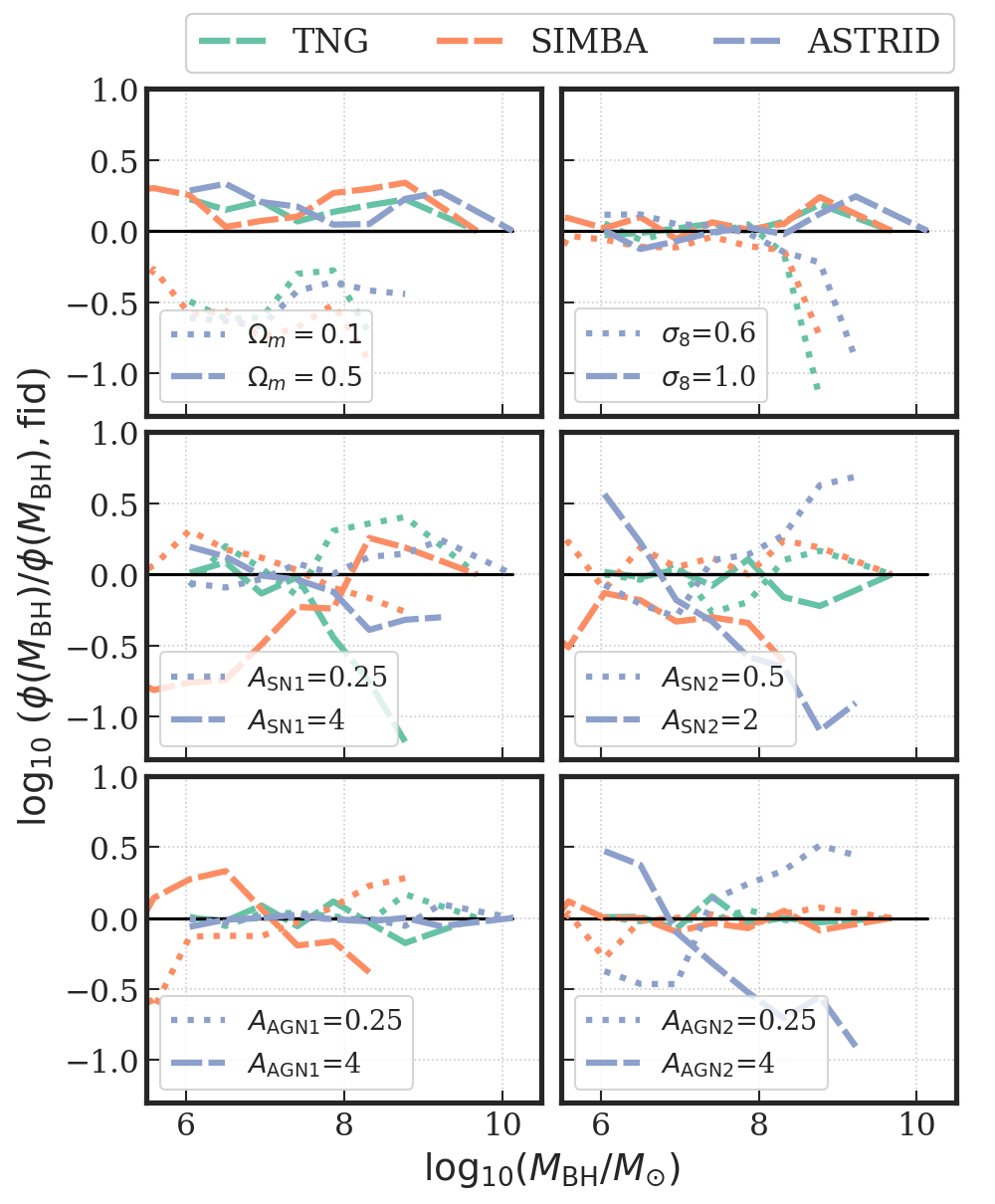}

\caption{
Impact of cosmological and astrophysical parameters on the galaxy stellar mass function (GSMF, the left panel) and black hole mass function (BHMF, the right panel). Both mass functions are evaluated at $z=0$. The $y$ axis gives the ratio between the mass function of the 1P simulation (with only 1 parameter varied and other parameters fixed) and the mass function of the fiducial run. The convention of colors and line styles is the same as that in the left panel Fig.~\ref{fig:1P-ps-sfrd}.
}
\label{fig:1P-GSMF-BHMF}
\end{figure*}


In this section, we diagnose the effect of the 6 fiducial cosmological and astrophysical parameters on a selection of the key physical quantities and compare the three simulation suites. 
In particular, we use the 1P set of the ASTRID, TNG, and SIMBA suites to investigate how varying one parameter (with other parameters fixed) would change the matter power spectrum, the global star formation rate history, as well as the galaxy and black hole populations.
Note that all the simulations in the 1P set have the same initial conditions, and therefore diminish the effects of cosmic variance in the comparison. 

We note that the parameter study in this section focuses on the 6 fiducial parameters ($\Omega_m$, $\sigma_8$, $A_{\rm SN1}$, $A_{\rm SN2}$, $A_{\rm AGN1}$, $A_{\rm AGN2}$) that are varied in the LH sets of ASTRID, TNG, and SIMBA.
In the Appendix, we show the effect of 22 additional parameters varied in the TNG and SIMBA through the extended 1P sets of TNG-1P-28 and SIMBA-1P-28.

\subsection{Matter power spectrum}
\label{subsetion: 1P-matter-power}

The left panels of Fig.~\ref{fig:1P-ps-sfrd} show the response of the matter power spectrum to the variation of the cosmological and astrophysical parameters individually. 
The fiducial model represents a simulation run with parameters \{$\Omega_m$, $\sigma_8$, $A_{\rm SN1}$, $A_{\rm SN2}$, $A_{\rm AGN1}$, $A_{\rm AGN2}$\} = 
 \{0.3, 0.8, 1, 1, 1, 1\}.

As expected, one can see that the matter power spectrum on large scales (with $k \lesssim 1 \invhmpc$) is mainly affected by the variation of cosmological parameters ($\Omega_m$ and $\sigma_8$), while the impact of the astrophysical parameters is limited to smaller scales of $k > 1 \invhmpc$ (with the exception of SIMBA, where the astrophysical parameters can mildly impact large scales due to long-range AGN jet feedback). 

We note that with the same variation in cosmological parameters, the matter power spectrum responds similarly on large scales but differently on small scales between the three simulation suites. 
For example, with increased $\Omega_m$ or $\sigma_8$, ASTRID predicts less relative enhancement in the small-scale matter power compared to TNG and SIMBA.  
This is due to the different astrophysical models in each simulation suite that interact differently with the varied cosmological parameters. 
Notably, compared to a smaller $\Omega_m$ (or $\sigma_8$), a larger $\Omega_m$ (or $\sigma_8$) promotes the (earlier) formation of more massive halos, galaxies and black holes, bringing non-linear effects on the small-scale matter power into play earlier, and resulting in the different small scale $P(k)$ variations among the three simulation models.
The interaction between cosmological parameters and astrophysical processes again stresses the importance of including different astrophysical models to improve the robustness of cosmological inference.

Variations in all four astrophysical parameters affect the amplitude and shape of the matter power spectrum in a non-trivial way.
In ASTRID, $A_{\rm AGN1}$ exhibits the weakest influence among all four astrophysical parameters. 
This is mainly due to the limited volumes of the simulations, which do not form the massive objects containing the most massive black holes that can trigger AGN jet feedback. 
The same trend is also seen in TNG and SIMBA, as all three simulation suites adopt a $M_{\rm BH}$-dependent AGN jet feedback model that biases jet feedback to more massive black holes. 
Given a larger volume, we might expect to see a more prominent impact of variations in AGN jet feedback on both galaxy formation and matter power spectrum.

The influence of SN and AGN feedback parameters is different between the 3 simulation suites, as they do not necessarily share the same physical meanings (as noted in Table~\ref{tab:table1}). 
Even for $A_{\rm SN2}$,  which represents the speed of hydrodynamically-decoupled galactic winds in the three simulation suites, its impact on the matter power spectrum (and on galaxy formation, as will be discussed later) can be quite different due to other `nuisance' parameters in detailed model implementations, the interplay with other astrophysical processes, and the non-linear evolution of galaxy formation.

It is expected that AGN feedback should make the dominant contribution to the baryonic suppression of the total matter power spectrum. 
Therefore, increased AGN feedback should lead to more suppression of the matter power.
However, we note that the total amount of AGN feedback cannot be simply controlled by the $A_{\rm AGN}$ feedback parameters, as the growth of galaxies and black holes are strongly regulated by feedback.

One example is the behavior of the $A_{\rm AGN2}$ parameter in the ASTRID suite. 
$A_{\rm AGN2}$ in ASTRID controls the efficiency of AGN thermal feedback, and we can see that a larger $A_{\rm AGN2}$ enhances (rather than suppresses) the matter power spectrum on small scales. 
As will be shown in Fig.~\ref{fig:1P-GSMF-BHMF}, this is because strong AGN thermal feedback aggressively suppresses the growth of the black hole population and ends up curtailing the total amount of injected AGN feedback energy. 
In particular, by limiting the formation of the massive black holes that are able to turn on jet mode feedback, strong AGN thermal feedback prevents an efficient suppression of the matter power. 
Therefore, we note that varying the feedback efficiency induces non-linear effects in the galaxy and black hole populations and eventually the matter power spectrum.

\subsection{Star formation rate density}
\label{subsetion: 1P-sfrd}

The right panel of Fig.~\ref{fig:1P-ps-sfrd} shows the change in the star formation rate density due to individual variations of the 6 cosmological and astrophysical parameters. 
At high redshifts $z>2$, the SFRD is significantly affected by changes in $\Omega_m$, $\sigma_8$ and also $A_{\rm SN1}$, while at lower redshifts the SFRD begins to be noticeably impacted by the $A_{\rm SN2}$ parameter in all three simulation suites and by $A_{\rm AGN2}$ in ASTRID.

We note that the $A_{\rm SN2}$ parameter in the ASTRID suite drives larger variations in star formation (also at higher redshifts) compared to TNG and SIMBA. 
$A_{\rm SN2}$ modulates the speed of hydrodynamically-decoupled galactic winds in all three simulation suites. 
However, the detailed SN wind models and the fiducial parameters are different.
Star formation in the ASTRID model turns out to be more sensitive to the SN wind speed 
than in TNG and SIMBA.
This behavior leads to a broader variation in the galaxy population in the ASTRID-LH set compared to the LH sets of TNG and SIMBA.

The left bottom panels of Fig.~\ref{fig:1P-ps-sfrd} indicate that the variation in AGN jet mode feedback ($A_{\rm AGN1}$ in all simulations and $A_{\rm AGN2}$ in TNG and SIMBA) has a limited impact on the global star formation. 
This result is somewhat affected by the small simulation volume, which limits the formation of the most massive systems that are subject to AGN jet feedback. 
The contribution of such systems to the global SFRD is however sub-dominant even without AGN feedback, except at late times, $z\lesssim1$ \citep{SpringelV_03b}.
We can see that the SFRD in SIMBA exhibits a relatively larger response to $A_{\rm AGN1}$ compared to ASTRID and TNG at lower redshifts ($z<2$). 
That is because the AGN jet mode in SIMBA hinges on a lower $\mbh$ threshold ($M_{\rm BH} > 10^{7.5} \msun$) compared to TNG and ASTRID, and therefore allows an earlier onset of AGN jet feedback that can impact the star formation rate.

Interestingly, we find that the AGN thermal mode feedback ($A_{\rm AGN2}$ in the ASTRID suite) begins to impact the global star formation at low redshift ($z<2$).
The enhanced AGN thermal feedback efficiency boosts global star formation.
This is partly a consequence of the non-linear impact of AGN feedback and self-regulation of black hole growth as discussed in Section~\ref{subsetion: 1P-matter-power}, where a higher $A_{\rm AGN2}$ efficiency limits the growth of black holes and eventually curtails the total budget of AGN feedback energy deposited into the interstellar medium.
We note that this effect is also seen in the TNG model when varying the AGN thermal feedback efficiency in the extended parameter study of TNG-1P-28, as shown in Appendix~\ref{sec:appendix-TNG-1P} (parameter 25, BlackHoleFeedbackFactor).


\subsection{Galaxy and black hole population at z=0}
\label{subsetion: 1P-gsmf-bhmf}

We now study the impact of the variation of the 6 parameters on the mass functions of the galaxy and black hole populations.
To investigate the variation across the mass range, Fig.~\ref{fig:1P-GSMF-BHMF} shows the ratio of the galaxy stellar mass function and black hole mass function at $z=0$ for each parameter variation compared to the fiducial model in the 1P set.

Fig.~\ref{fig:1P-GSMF-BHMF} shows that the galaxy and black hole populations respond to $\Omega_m$ in all mass intervals, while $\sigma_8$ has less impact compared to $\Omega_m$ and only affects the galaxy and black hole populations at the most massive end at $z=0$. 
As it sets the amplitude of the initial density fluctuations, $\sigma_8$ modulates the onset of gravitational collapse and structure formation. 
Due to hierarchical structure formation, in which small galaxies form earlier and merge to form more massive galaxies later, a lower $\sigma_8$ with delayed structure formation leaves an imprint on the abundance of massive galaxies, 
while the population of less massive galaxies has sufficient time evolution to grow down to $z=0$.
The impact of $\sigma_8$ is thus larger at higher redshifts. 
For example, at $z=4$, the abundance of $M_{*} \sim 10^{9} \msun$ galaxies can be $0.5$ dex higher (lower) with $\sigma_8 = 1.0$ (0.6).  
Therefore, in general, it is more challenging to do parameter inference for $\sigma_8$ compared to $\Omega_m$ based on halo or galaxy populations at $z=0$ in CAMELS \citep[e.g., see][]{Helen_2022,deSanti2023}, due to the limited number of massive systems in the simulation volume.

The impact of the $A_{\rm SN1}$ parameter on the galaxy population exhibits a similar trend among the three simulation suites, with larger SN feedback efficiency leading to suppression of the galaxy abundance in all mass intervals.
For the black hole population, we note that a larger $A_{\rm SN1}$ greatly suppresses the low mass black hole population in SIMBA, as SIMBA adopts a black hole seeding prescription based on the stellar mass threshold $M_* > 10^{9.5} \msun$ \citep{Dave2019MNRAS.486.2827D,Thomas2019,Habouzit2020}, which is affected by SN feedback (while the TNG and ASTRID suites seed black holes based only on halo mass).

The effects of $A_{\rm SN2}$ in the ASTRID suite are more drastic compared to TNG and SIMBA, where a larger $A_{\rm SN2}$ significantly reduces the abundance of galaxies (especially at the massive end), as well as the population of massive black holes, as black holes accrete from the gas reservoir that also fuels star formation. 
We note that due to the halo-based black hole seeding model in CAMELS-ASTRID, the total number of black holes seeded is not affected by the astrophysical parameters. Therefore, in the cases where black holes do not grow efficiently, the black hole population accumulates at the low mass end.

The $A_{\rm AGN1}$ parameter, controlling the energy (in TNG and ASTRID) or the momentum flux (in SIMBA) of the AGN jet feedback, exhibits limited impact on the galaxy and black hole populations, with the black hole population in SIMBA showing a larger response to $A_{\rm AGN1}$ as the jet mode feedback hinges on a lower $\mbh$ threshold and can regulate black hole growth at an earlier stage, as previously discussed in Section~\ref{subsetion: 1P-sfrd}.

The $A_{\rm AGN2}$ parameter controls the black hole thermal feedback efficiency in the CAMELS-ASTRID suite (as opposed to the jet speed in TNG and SIMBA).
As the AGN thermal feedback is the dominant feedback channel that strongly regulates the growth of black holes before they become massive enough to turn on AGN jet feedback, $A_{\rm AGN2}$ in ASTRID shows a prominent impact on both the black hole and galaxy populations. 
As discussed in earlier sections, a larger $A_{\rm AGN2}$ suppresses the formation of massive black holes, thereby curtailing AGN feedback on galaxy formation and producing an enhancement of the global star formation rate and galaxy populations in the ASTRID suite.

\begin{figure*}
\centering
\includegraphics[width=0.98\textwidth]{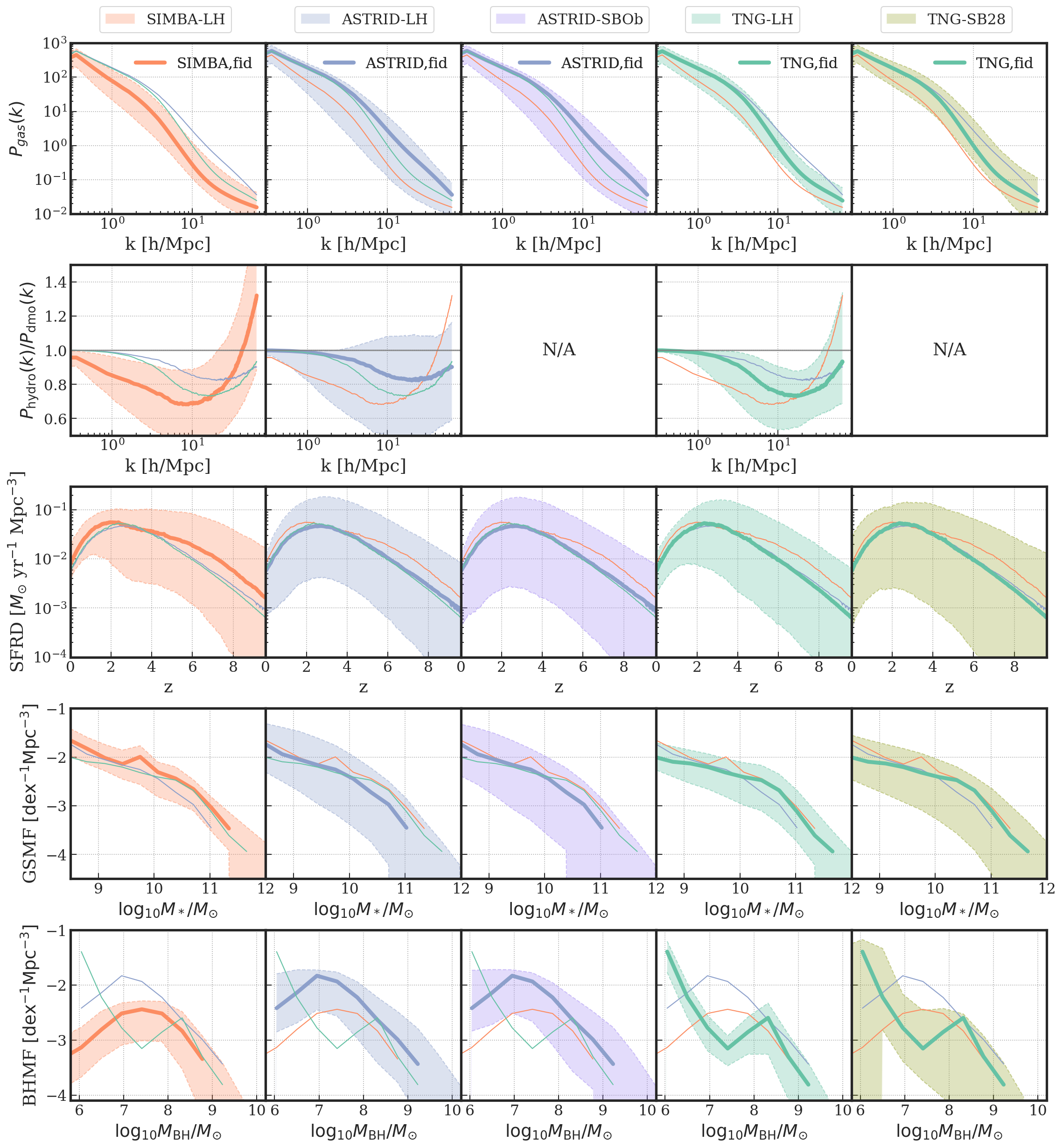}
\caption{Illustration of the range of different cosmological and astrophysical properties in the 5 large simulation suites: SIMBA-LH (orange), ASTRID-LH (blue), ASTRID-SBOb (purple), TNG-LH (green) and TNG-SB28 (yellow green). 
From top to bottom the panels show the gas power spectrum (1st), the ratio of the matter power spectrum in hydrodynamic simulations to that of the corresponding N-body simulations (2nd), the star formation rate density history (3rd), the stellar mass function (4th) and the black hole mass function (5th). 
The power spectra and mass functions are shown at $z=0$. Note that the halo masses and stellar masses are from the SUBFIND catalogs. 
For each panel, the orange, green and blue lines give the predictions from the fiducial models of SIMBA, TNG and ASTRID respectively, calculated by taking the median of the CV set. 
Each simulation suite has its corresponding fiducial model result emphasized in a thicker line. 
The shaded regions indicate 10-90 percentiles from each large LH simulation set (or SB set for ASTRID and TNG extension).
}
\label{fig:CV-LH-1D}
\end{figure*}



\section{Overview of the training set variation range}
\label{section: CVLH}

In this section, we present an overview of some key cosmological and astrophysical properties for the five  large hydrodynamic simulation training sets currently available in CAMELS: the three LH sets of TNG, SIMBA and ASTRID (1000 simulations each, with 6 varied parameters) and the two extension sets of ASTRID-SBOb and TNG-SB28 (1024 simulations each, with additional varied parameters). 
We note that many of these cosmological and astrophysical properties have been introduced in the first CAMELS presentation paper \citep{Villaescusa-Navarro2021ApJ...915...71V} for the TNG and SIMBA LH simulation sets.
Here we revisit some of those quantities by comparing the fiducial astrophysical models of ASTRID with TNG and SIMBA, and showing the range of variation of those properties covered by the 5 simulation sets.
We present several 1-dimensional statistics in Section~\ref{subsection: CVLH-1D} and various scaling relations between halos, gas, galaxies and black holes in Section~\ref{subsection: CVLH-galaxy-catlog}.


\subsection{1D statistics}
\label{subsection: CVLH-1D}

We first look at the 1D statistics of the matter power spectrum, star formation rate density, and mass functions as shown in Fig.~\ref{fig:CV-LH-1D}.
Note that for each of the panels in Fig.~\ref{fig:CV-LH-1D}, the solid lines give the median value in the CV sets from the fiducial astrophysical models of ASTRID, TNG and SIMBA, and the shaded regions give 10-90 percentiles in the LH (or SB) sets to illustrate the variation range of the property considered. 
All quantities are measured at $z=0$ except for the star formation rate density.
We describe each quantity in more detail below.

\textbf{Gas power spectrum}: 
The first row of Fig.~\ref{fig:CV-LH-1D} shows the power spectrum of the gas component $P_{\rm gas}(k)$, which exhibits systematic differences between the three hydrodynamic simulation models.
For the result of the CV set (the fiducial model), the clustering of gas is highest in ASTRID and is lowest in SIMBA, with the difference reaching one order of magnitude at the scale of $k \sim 10 \invhmpc$.
As already discussed in Section~\ref{section: CV-gas}, among the three hydrodynamic simulation models, the fiducial version of ASTRID has the mildest AGN feedback and therefore leads to the lowest suppression of gas clustering. 

In spite of this, the variation of the gas power in ASTRID is the largest among the simulation suites on small scales. 
At $k \sim 10 \invhmpc$, the gas power can vary across two orders of magnitudes in the ASTRID LH set, covering the total range of SIMBA and TNG, and even TNG-SB28 that varies a significantly larger number of model parameters than ASTRID-LH. 
This large variation in the small-scale gas power is driven by the combination of $A_{\rm AGN2}$ and the two $A_{\rm SN}$ parameters, as can be inferred from Fig.~\ref{fig:1P-ps-sfrd}.

The similarity between ASTRID-LH and ASTRID-SBOb, and further, even between TNG-SB28 and TNG-LH, where the addition of 22 varied parameters increases the diversity of $P_{\rm gas}(k)$ responses only by a modest factor, does not demonstrate that variations in $\Omega_b$ or any other of these additional parameters are necessarily sub-dominant for $P_{\rm gas}(k)$ with respect to the original 6 parameters of the LH sets. 
The $P_{\rm gas}(k)$ responses to the individual additional parameters are, in fact, in some cases no less significant than the original ones, as shown explicitly in Appendix \ref{sec:appendix-TNG-1P} and \ref{sec:appendix-SIMBA-1P}. 
Instead, we interpret the rough similarity between the diversity of $P_{\rm gas}(k)$ results in the original and extended parameter spaces to be a result of a `regression towards the mean' effect, whereby the variations in a larger number of relevant parameters tend to roughly cancel each other out. 
Nevertheless, by successfully using these simulation sets in future work to learn the relations between the parameters and simulation results, such as $P_{\rm gas}(k)$, we will be able to probe the entirety of these extended parameter spaces, including regions therein that are far from the sampling points directly simulated here. 
We do expect this task, however, to become more challenging, as the sparsity of this sampling increases significantly due to the `curse of dimensionality'.


\textbf{Matter power spectrum ratio}: 
We consider the total matter power spectra for each hydrodynamic simulation in comparison to the matter power spectrum of the corresponding dark matter only runs and give the value of $P_{\rm hydro}/P_{\rm dmo}$ in the second panel of Fig.~\ref{fig:CV-LH-1D}.
The $P_{\rm hydro}/P_{\rm dmo}$ ratio for the extension simulation sets of ASTRID-SBOb and TNG-SB28 is skipped due to the absence of the companion dark matter only simulations. 

Compared to Fig.~\ref{fig:PsRatio-CV}, where the shaded regions give the cosmic variance driven by different realizations in the CV set, we can see that the variation of $P_{\rm hydro}/P_{\rm dmo}$ in the LH sets is significantly larger. 
The ASTRID-LH set can yield larger variation compared to the LH set of SIMBA and TNG, which reflects the similar trend pointed out above for $P_{\rm gas}(k)$.
In some of the strong feedback scenarios, the power spectrum ratio of ASTRID can reach $\sim 50 \%$ on small scales ($k \sim 10 \invhmpc$).
On the other hand, some of the astrophysical models in the ASTRID-LH set can produce $P_{\rm hydro}/P_{\rm dmo}>1$, which is not seen in the TNG and SIMBA suites.
Simulations in those scenarios usually have a limited number of massive black holes 
(caused by, e.g. a large $A_{\rm AGN2}$ parameter), and the AGN kinetic (jet) feedback cannot efficiently counteract the small clustering from star formation.
Therefore, those simulations predict enhancement rather than suppression of the total matter power.



\textbf{Star formation rate density}: 
The third panel of Fig.~\ref{fig:CV-LH-1D} shows the global star formation rate density for each of the simulation sets. 
The shaded regions indicate the 10-90 percentiles for the given redshift bin in the LH (or SB) sets.
Comparing the three LH sets, the SFRDs in the ASTRID-LH set display larger variation at lower redshift $z < 2$ compared to the LH sets of SIMBA and TNG.
We can see from the right panel of Fig.~\ref{fig:1P-ps-sfrd} that the SFRD in ASTRID at low redshift is very sensitive to $A_{\rm SN2}$. 
The same level of variation in $A_{\rm SN2}$ in the ASTRID suite drives a larger diversity of SFRDs compared to TNG and SIMBA.
Furthermore, the SFRD at low redshifts is more sensitive to $A_{\rm AGN2}$ in ASTRID. 
Therefore, ASTRID-LH also features a larger variation in the star formation history, as well as the galaxy population at $z=0$ compared to the other two LH sets. 

With the additional parameters varied, the extension suites of ASTRID-SBOb and TNG-SB28 both show mildly larger variation in the SFRDs compared to their corresponding LH sets (more so for the larger extension represented by TNG-SB28 than that of ASTRID-SBOb, as might be expected). 
Our interpretation for this follows the ones we presented for $P_{\rm gas}(k)$ above.

\begin{figure*}
\centering
\includegraphics[width=1.0\textwidth]{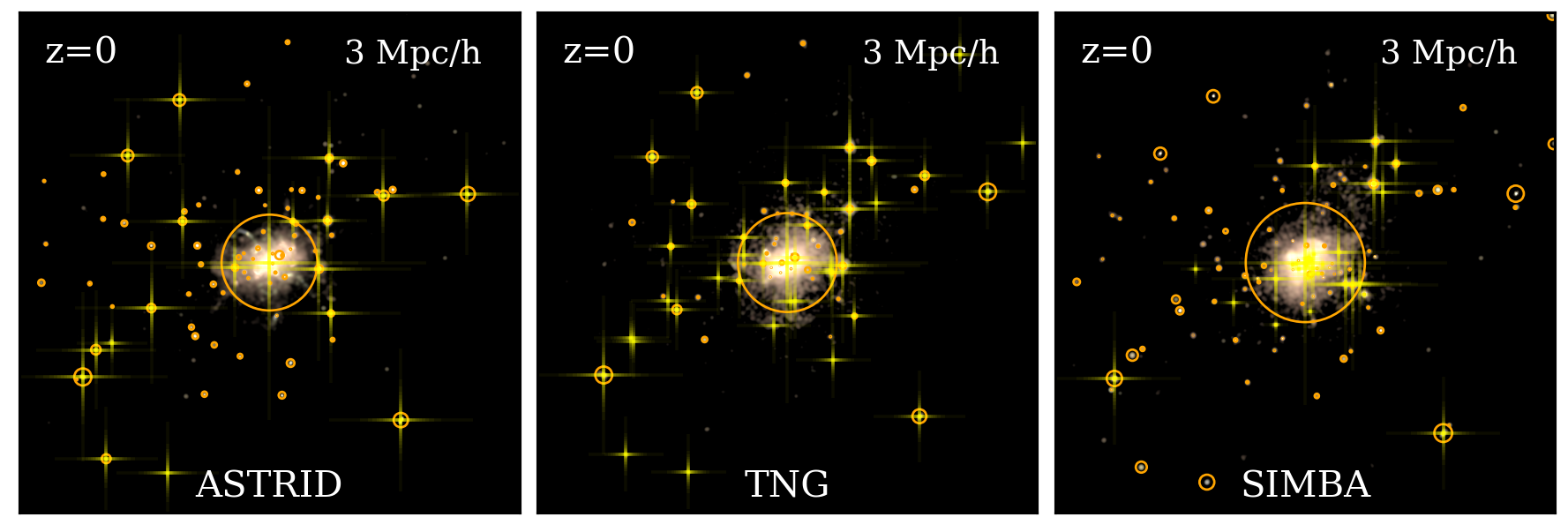}
\caption{Illustration of galaxies and black holes in the CV0 simulation for ASTRID (first column), TNG (second column), and SIMBA (third column). 
Each panel shows the stellar density field over a region of $(3 \hmpc)^3$ centered on the most massive halo in the simulation.
The orange circles mark galaxies with stellar mass $M_{*} > 10^{8} \msun$, as identified by SUBFIND. The radii of the circles correspond to the virial radii of the subhalos. 
The yellow spikes mark the positions of all black holes in this region, with the size scaled by $M_{\rm BH}$.}
\label{fig:StarImage}
\end{figure*}

\textbf{Galaxy and black hole population}: 
The bottom two rows of Fig.~\ref{fig:CV-LH-1D} present the galaxy stellar mass function (GSMF) and black hole mass function (BHMF) at $z=0$.
The stellar masses are summary statistics from the subhalo catalogs given by the SUBFIND algorithm, including the contribution of both host and satellite halos (galaxies). 
The statistics of galaxies and black holes exhibit systematic differences among the simulation suites due to the different astrophysical models. 

Fig.~\ref{fig:StarImage} illustrates the galaxy and black hole population at $z=0$ in the three fiducial models of ASTRID, TNG and SIMBA over a projected region of $(3 \hmpc)^3$ surrounding the most massive halo in the CV0 simulations. 
The orange circles mark the positions of galaxies with $\mstar>10^{8} \msun$ and the yellow spikes mark the locations of all black holes. 
The radii of circles and sizes of the spikes are scaled by $M_{*}$ and $M_{\rm BH}$ respectively. 
The large-scale pattern of the galaxies is similar among the three simulations due to the same initial conditions and therefore the resulting large-scale structure.
Meanwhile, we can see differences in the population of small galaxies and black holes due to the various astrophysical models used in the three hydrodynamical simulations. 
Noticeably, the small satellite galaxy population is suppressed in TNG compared to ASTRID and SIMBA, as also indicated by the GSMF in Fig.~\ref{fig:CV-LH-1D}.
On the other hand, small galaxies in SIMBA do not host black holes due to its black hole seeding model. 
Therefore, we can see that SIMBA produces a smaller black hole population as well as lower occupation fraction in galaxies compared to the other two simulations.

For the fiducial models, galaxy abundance in ASTRID is slightly lower than in TNG and SIMBA by $\sim 0.3$ dex in the stellar mass interval $M_{*} = 10^{10 - 11} \msun$. 
SIMBA features a spike in the galaxy mass around $M_{*} = 5\times10^{9} \msun$. 
The TNG galaxy abundance is slightly lower at the low mass end of $M_{*} < 10^{9} \msun$ compared to the other two models.

The variation of the GSMF in the ASTRID-LH set is larger than the LH sets of TNG and SIMBA for the same reasons as discussed for the SFRD. 
At the low mass end ($M_{*} < 10^{10} \msun$), the range of variation in ASTRID covers that of the SIMBA and TNG simulation suites, including TNG-SB28. 
The broad variation of the galaxy population in ASTRID-LH allows machine-learning models trained on the ASTRID galaxy catalogs (of the LH set) to provide the best extrapolation results when applied to other simulation suites as test sets \citep[e.g.][]{deSanti2023, Nicolas-prep}.

Compared to the galaxy population, the black hole mass function shows much larger differences between the fiducial models of the three simulation suites, especially at the low mass end of $\mbh < 10^{8} \msun$, due to different prescriptions for black hole seeding, accretion and feedback (as well as poor observational constraints for model calibration; see also \citealt{Habouzit2020}). 
Compared to TNG and SIMBA, ASTRID has a larger black hole population especially at the low mass end of $M_{\rm BH} = 10^{6 - 8} \msun$.
The low-mass end black hole population can be sensitive to the black hole seed mass.
We reiterate that the CAMELS-ASTRID suite does not adopt the same black hole seeding prescription as the large volume ASTRID production run due to the limited resolution of CAMELS, as discussed in Section~\ref{sec:method}.

The diversity of the black hole mass function as a result of parameter variations is larger in ASTRID than in the LH sets of the other suites, similarly to the case for the quantities discussed above. 
It is also true that the extension into $\Omega_b$ variations in ASTRID-SBOb does not noticeably increase the diversity compared to ASTRID-LH. 
However, in the case of the black hole mass function, TNG-SB28 displays a much larger diversity than TNG-LH, or even than the ASTRID sets, unlike what was seen above for other quantities. 
We interpret this as being a result of a high sensitivity of black hole masses to a number of the new parameters introduced in TNG-SB28, which include specifically a significant number of parameters that directly control black hole growth. 
This is as opposed to the 6 parameters in the original TNG-LH set, most of which made little difference for the growth of black holes.

\begin{figure*}
\centering
\includegraphics[width=1.0\textwidth]{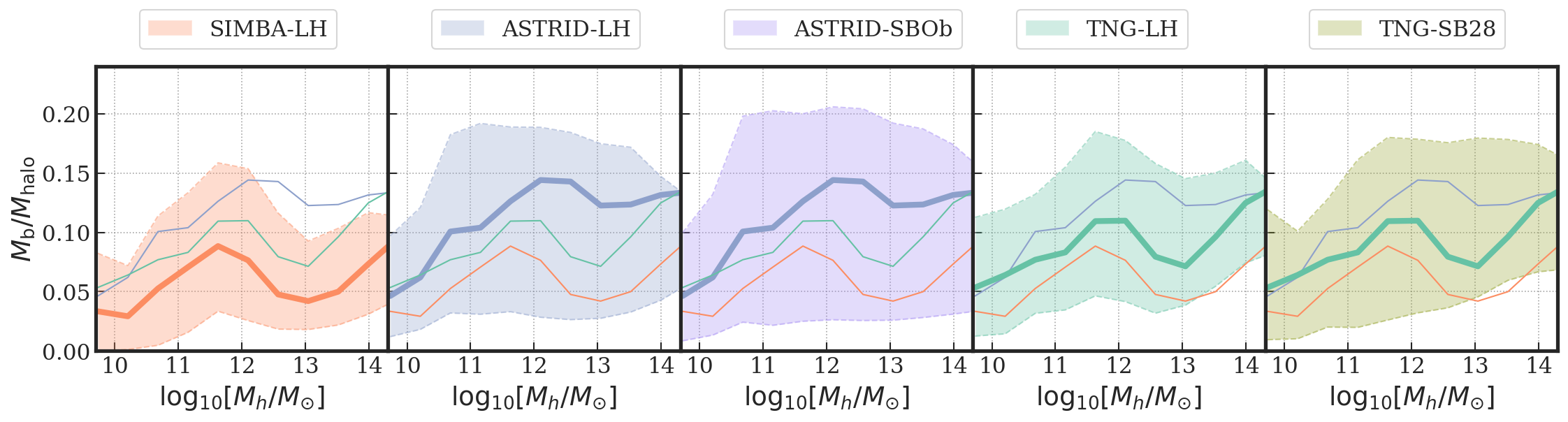}
\includegraphics[width=1.0\textwidth]{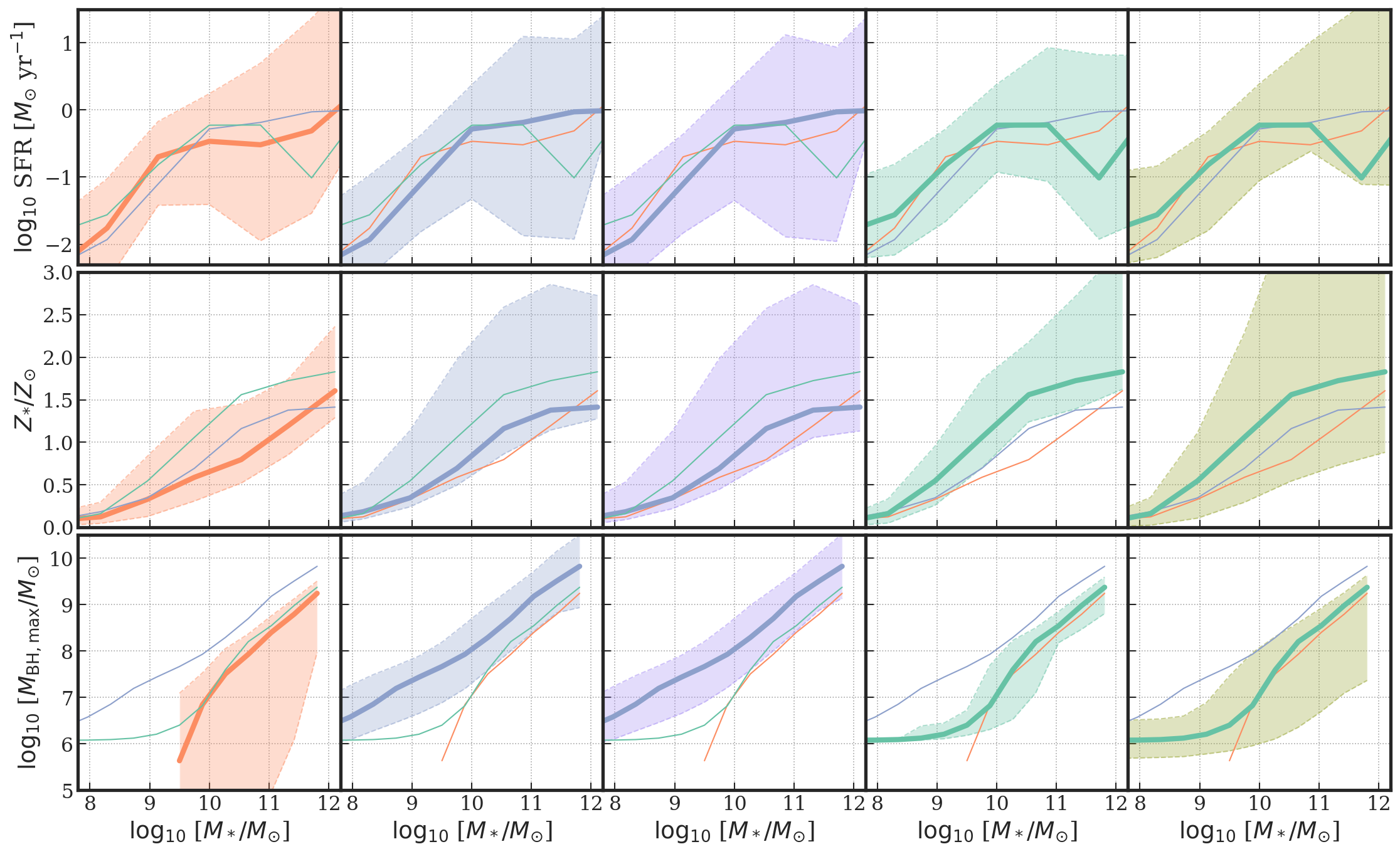}
\caption{
Various scaling relations between astrophysical properties from the (sub)halo catalogs at $z=0$. Conventions for the color, lines and shaded regions are the same as in Fig.~\ref{fig:CV-LH-1D}.
The top panel shows the halo baryon fraction versus halo mass.
The lower three panels show the scaling relations between stellar mass and star formation rate (2nd), mean stellar metallicity (3rd) and the maximum $M_{\rm BH}$ (4th).
For each property, we combine the subhalo catalogs from the CV set and LH set and show in solid lines the median value of $y$ for each mass bin in $x$ from the CV set, and in the shaded area the 10-90 percentile of the $y$ distribution from the combined catalog from 1000 simulations of the LH set (or 1024 simulations of the SB set). See text for more details.}
\label{fig:CV-LH-scaling}
\end{figure*}


\subsection{Galaxy catalogs and scaling relations}
\label{subsection: CVLH-galaxy-catlog}

In this section, we inspect the halo and galaxy catalogs in the 5 simulation sets and compare various scaling relations of astrophysical properties.

\textbf{Halo baryon fraction}:
The first row of Fig.~\ref{fig:CV-LH-scaling} shows the scaling relation between the baryon fraction $f_b = M_{\rm b}/M_{\rm halo}$ and $M_{\rm halo}$, where $M_{\rm halo}$ here is the FOF halo mass, and $M_{\rm b}$ is the total baryonic mass (including the stars, gas and black holes) in the FOF halo.
We can see the systematic difference between the three simulation suites.
The baryon fraction in the three simulation suites all peak at around $M_{h} \sim 10^{12} \msun$ and decrease when going to larger halo mass as the result of AGN (jet) feedback in massive halos, where ASTRID exhibits the largest baryon fraction for massive halos while SIMBA predicts the lowest.
A smaller baryon fraction in halos corresponds to less clustering of gas \citep[see, e.g.][for more detailed studies]{vanDaalen2020,Delgado2021,Delgado2023,Pandey2023arXiv230102186P}.
Therefore, the baryon fraction in the three simulation suites exhibits a similar trend as the gas power spectrum as seen in Fig.~\ref{fig:CV-LH-1D}. 
Meanwhile, the ASTRID-LH set displays the largest variation in the baryon fraction within the same halo mass bin, almost covering the range of TNG and SIMBA, whether or not the LH or SB sets are considered.

\textbf{Star formation rate versus stellar mass}:
The second row of Fig.~\ref{fig:CV-LH-scaling} shows the galaxy star formation rate (SFR) versus the stellar mass of galaxies from the galaxy subhalo catalogs. 
Here SFR is the instantaneous star formation of all gas particles associated with their galaxy hosts.
We only consider galaxies with non-zero star formation rates to avoid the distributions being heavily affected by fully quenched galaxies.
The three simulation suites exhibit a similar trend in that the SFR increases with stellar mass and flattens when going to more massive galaxies due to quenching from feedback. 
As discussed in \cite{Villaescusa-Navarro2021ApJ...915...71V}, in the low $M_*$ end, cosmic variance can make the dominant contribution to the variation in the SFR-$M_*$ relation.
Therefore, we see a similar range of the LH set variation for the three simulation suites for lower mass galaxies at $M_{*} < 10^{10} \msun$. 
The variation of the SFR increases at higher galaxy masses.
The SFR variation can exceed 3 orders of magnitude for $M_* > 10^{11} \msun$ in the three simulation suites, as the AGN feedback begins to quench the galaxies, therefore giving rise to a larger scatter in this mass regime by variations of the feedback parameters in the LH and SB sets.

\textbf{Metallicity versus stellar mass}:
The third row of Fig.~\ref{fig:CV-LH-scaling} shows the relation between stellar metallicity, in units of solar metallicity ($Z_{\odot}=0.012$), and stellar mass. The three simulation suites exhibit systematically different $Z_* - M_*$ relations (or metal enrichment histories) due to different astrophysical models.
The fiducial model of TNG predicts the largest stellar metallicity in all stellar mass bins, while ASTRID gives $Z_*$ between SIMBA and TNG. 
The variation in the ASTRID-LH set tends to be larger than the other two LH sets, especially for the massive galaxies, but TNG-SB28 shows a much higher diversity than both TNG-LH and ASTRID-LH. We believe this is due to the particularly large effect of two of the additional parameters in TNG-SB28 on this relation, specifically the slope of the IMF and the metal loading factor of the galactic winds.

\textbf{Black hole mass versus stellar mass}:
The bottom panel of Fig.~\ref{fig:CV-LH-scaling} shows the $M_{\rm BH} - M_{*}$ relation.
For each galaxy identified by SUBFIND, we consider the single most massive black hole associated with that galaxy.
We only include galaxies with non-zero black holes, which results in the cut-off at $M_{*} \lesssim 10^{9.5} \msun$ in SIMBA due is its $M_*$ threshold for black hole seeding.

$M_{\rm BH}$ in the ASTRID suite is systematically larger than TNG and SIMBA within a given stellar mass bin. 
For lower mass galaxies $M_* < 10^{10} \msun$, ASTRID predicts $M_{\rm BH}$ about 0.5-1 dex more massive compared to TNG, as black holes in ASTRID grow more rapidly in the initial phase with boosted Bondi-accretion factor $\alpha=100$ and the possibility of super-Eddington accretion (the $\dot{M}_{\rm BH}$ in ASTRID is capped at $2\times L_{\rm Edd}$). 
At higher $M_{\rm BH}$ (or $M_*$), black hole growth starts to become self-regulated by AGN feedback.
At higher masses $M_* > 10^{10} \msun$, ASTRID galaxies host $M_{\rm BH}$ about 0.4 dex more massive compared to TNG and SIMBA.
We note again that the $M_{\rm BH}$ at $z=0$ as well as the $M_{\rm BH} - M_*$ relation can be sensitive to the initial black hole seed mass, where the CAMELS-ASTRID suite applies a different black hole seeding prescription compared to the ASTRID production run because of the lower resolution of CAMELS.

The range of variation in the resulting $M_{\rm BH}$-$M_*$ relations when the simulation parameters are varied is quite distinct between the five different sets, with the exception of ASTRID-LH and ASTRID-SBOb, which are virtually the same. 
In particular, this diversity is a strong function of stellar mass, with different dependences on mass in the different sets. 
We see again, similarly to the black hole mass function, that the diversity in TNG-SB28 is much larger than in TNG-LH, which we associate with the addition of new parameters that directly affect black hole growth and evolution.

\begin{figure*}
\centering
\includegraphics[width=0.9\textwidth]{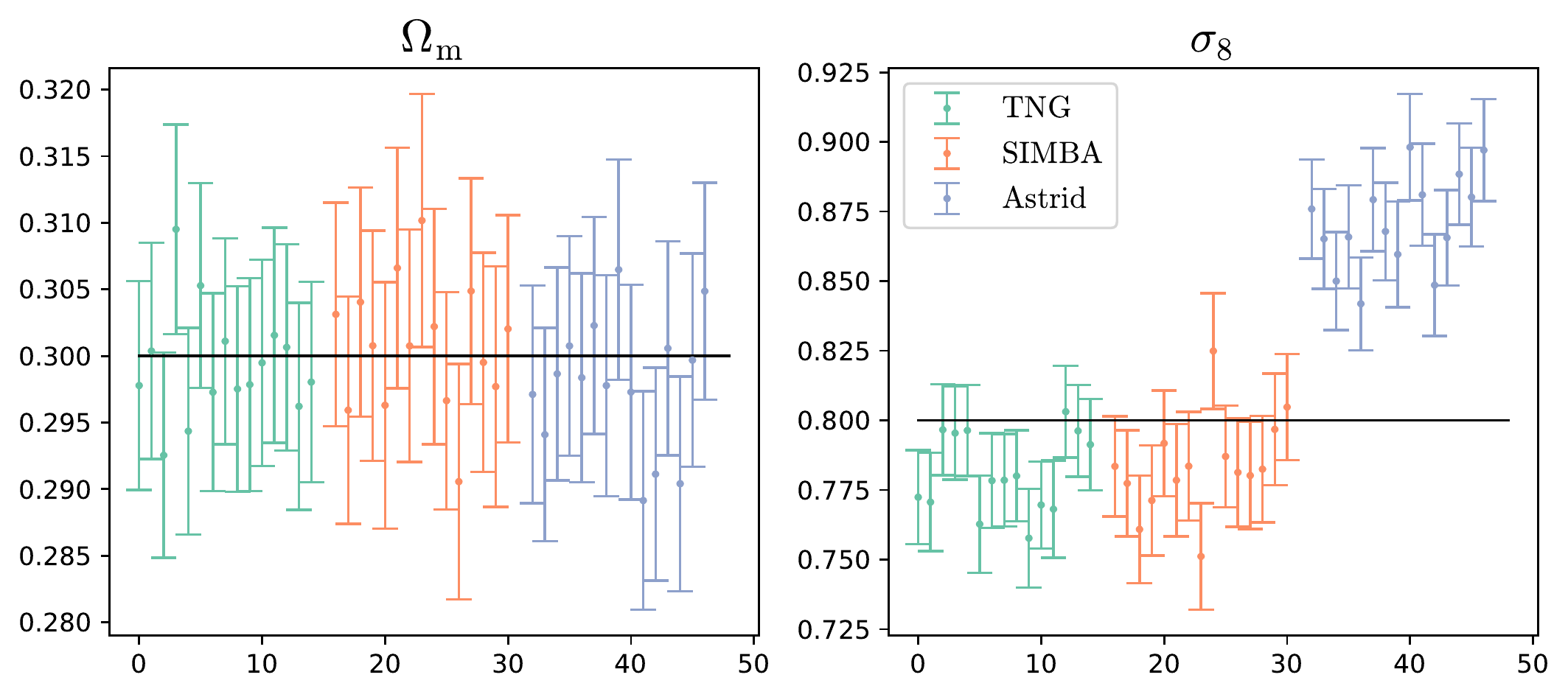}
\caption{We use the model described in \citet{Paco_robust_maps}, which was trained on total matter maps from CAMELS-TNG simulations, to conduct tests on 15 distinct maps with constant values of $\Omega_{\rm m}$ and $\sigma_8$ (shown with black horizontal lines) from three different simulation suites: TNG (green), SIMBA (orange), and ASTRID (blue). As observed, the model performed effectively in inferring the value of $\Omega_{\rm m}$, however, it was not able to correctly infer $\sigma_8$ from CAMELS-ASTRID maps.}
\label{fig:Mtot_inference}
\end{figure*}

\begin{figure*}
\centering
\includegraphics[width=0.9\textwidth]{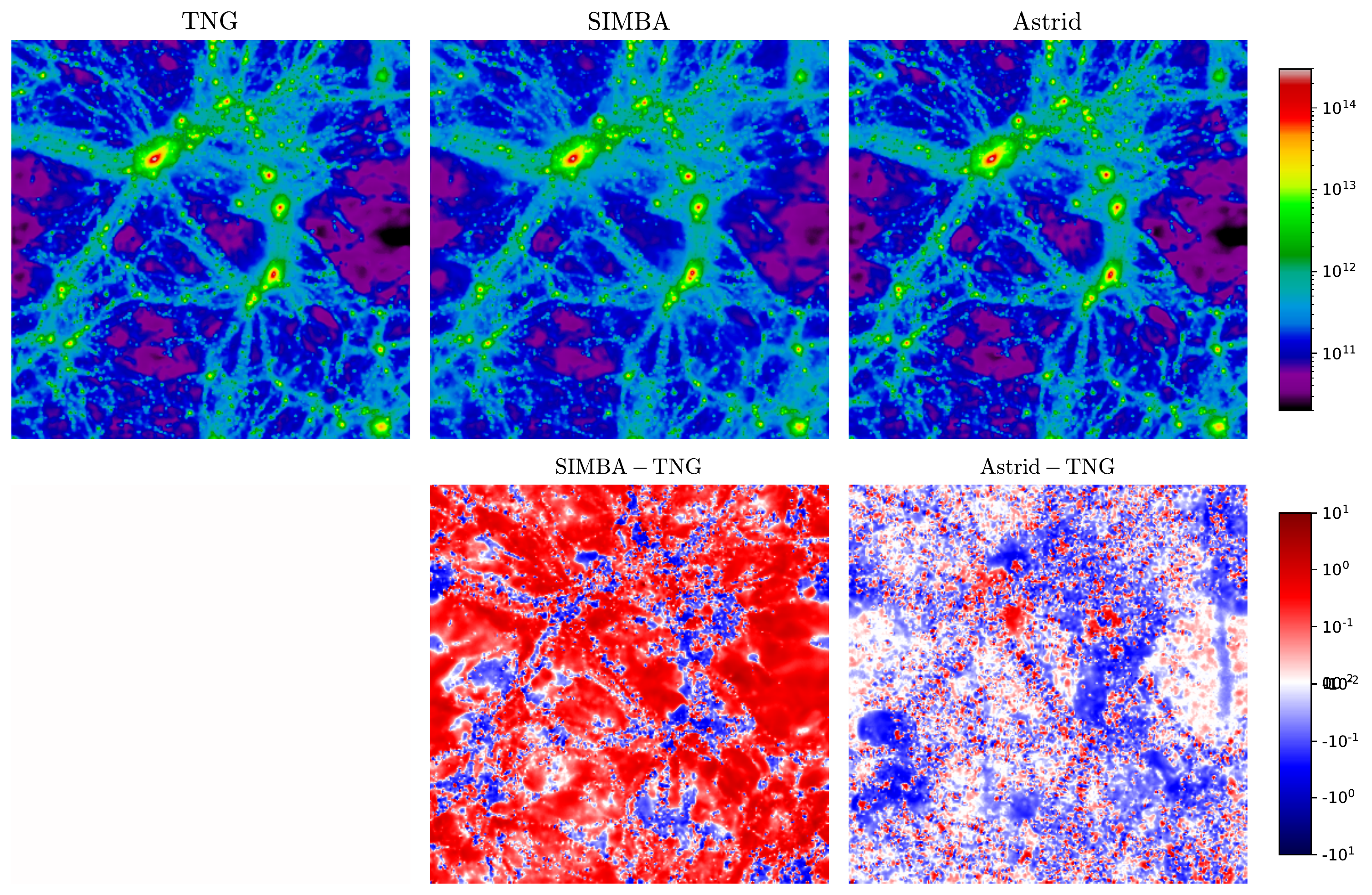}
\caption{The top row shows three images with the total matter surface density from three different simulations run with the same initial conditions. The panels in the bottom row show differences with respect to the TNG image. From the residual plots, it can be seen that the images from the TNG and ASTRID models are more similar to the ones from SIMBA.}
\label{fig:Mtot_maps}
\end{figure*}


\section{Machine learning tests}
\label{Sec:tests}

In this section, we demonstrate several machine learning applications that can be performed using the new simulation suites introduced in this work. 
As a simple demonstration, we carry out ML tasks following our previous work on field-level likelihood-free inference of $\Omega_m$ and $\sigma_8$ using 2D maps of the total matter density field and gas temperature field.
These 2D maps represent the projection of $25\times25\times5~(h^{-1}{\rm Mpc})^3$ slices of the simulation volume and have been created for the new CAMELS-ASTRID and TNG-SB28 simulations using the same procedure as presented in \cite{CMD}. 
These maps will be included in the CAMELS Multifield Dataset\footnote{\url{https://camels-multifield-dataset.readthedocs.io}}.

Section~\ref{subsec6.1:robustness} tests previously developed ML models that are robust across the two subgrid models of TNG and SIMBA on the new ASTRID field. 
In Section~\ref{subsec6.2:two-suite}, we demonstrate the case of training ML models based on a combination of two simulation suites (LH sets of TNG + SIMBA) and testing them on the ASTRID LH set. 
In Section~\ref{subsec6.3:TNG28-tests}, we present the first use of the new TNG-SB28 suite as a training set to showcase the possibility of developing ML inference models based on a simulation set embedded in a high-dimensional parameter space.

We note that in this section, we use the new CAMELS-ASTRID suite to test model robustness and validate the performance of ML models trained on two simulation suites.  
ML models \textit{trained} on the new ASTRID suite have already been presented in \cite{Pandey2023arXiv230102186P, Helen_2022, deSanti2023} and have exhibited good performance. 
In particular, \cite{deSanti2023} showed that, among all the available suites, ASTRID was the one with the best generalization properties when studying galaxy clustering, while \cite{Nicolas-prep} reached similar conclusions when studying properties of individual galaxies.

\subsection{Robustness tests}
\label{subsec6.1:robustness}

In this section, we investigate whether models that are robust across two simulation suites will also be accurate when tested on simulations from a third suite.
For this, we take advantage of the results presented in \citet{Paco_robust_maps} where it was shown that a convolutional neural network (CNN) model was able to perform a robust field-level likelihood-free inference on the values of $\Omega_{\rm m}$ and $\sigma_8$ from 2D maps containing slices of the total matter density field.

The network was trained on maps created from the LH set of TNG simulations and tested on maps from both TNG and SIMBA, showing that the approach was able to infer the values of $\Omega_{\rm m}$ and $\sigma_8$ with comparable accuracy and precision. 
The same test was performed training the model on SIMBA maps yielding the same conclusion. 
Having a model that is robust across the TNG and SIMBA models, we investigate whether it is also robust when tested on data from ASTRID.

The result we find is that the TNG-trained model is able to infer the true value of $\Omega_{\rm m}$ from the maps of all three simulations, while for $\sigma_8$ it is only able to perform such a task in the TNG and SIMBA maps, but fails on the maps from ASTRID. This is demonstrated in Fig.~\ref{fig:Mtot_inference}, which shows the results of the model trained on TNG and tested on the CV set of the three simulation suites (so the true values of $\Omega_m$ and $\sigma_8$ are 0.3 and 0.8 respectively).

To illustrate the model performance on the three simulation suites,
Fig.~\ref{fig:Mtot_maps} gives an example of the total mass map from the TNG, SIMBA and ASTRID CV set (with the same realization) as one of the test sets. From the visual comparison as well as the residual plots shown in the bottom panels of Fig.~\ref{fig:Mtot_maps}, one can see that the total matter maps of TNG and ASTRID are more similar to each other compared to the one from SIMBA. 
However, the network that can infer $\sigma_8$ for the TNG map (and for the SIMBA map) gives a biased result when tested on the ASTRID map. 
We also test the network on the CV set of the Magneticum simulation (that also contains 27 CV simulations) and find that the model also gives a biased result of $\sigma_8$ similar to that of ASTRID. 
We speculate that the network infers $\sigma_8$ contingent on some complex, high-order (and possibly numerical) features that happen to be shared between TNG and SIMBA while broken by ASTRID (and Magneticum) maps.
We leave it to future work to interpret and understand such behavior and develop a more robust model for $\sigma_8$ inference. 

We stress that the falsification of the model on the $\sigma_8$ inference task demonstrates the importance of testing ML models on a range of independent galaxy formation simulations to ensure their robustness and generalizability, as ML models that can work on two galaxy formation simulations are not guaranteed to perform well when tested on a third one.

\begin{figure*}
    \centering
    \includegraphics[width=1.0\textwidth]{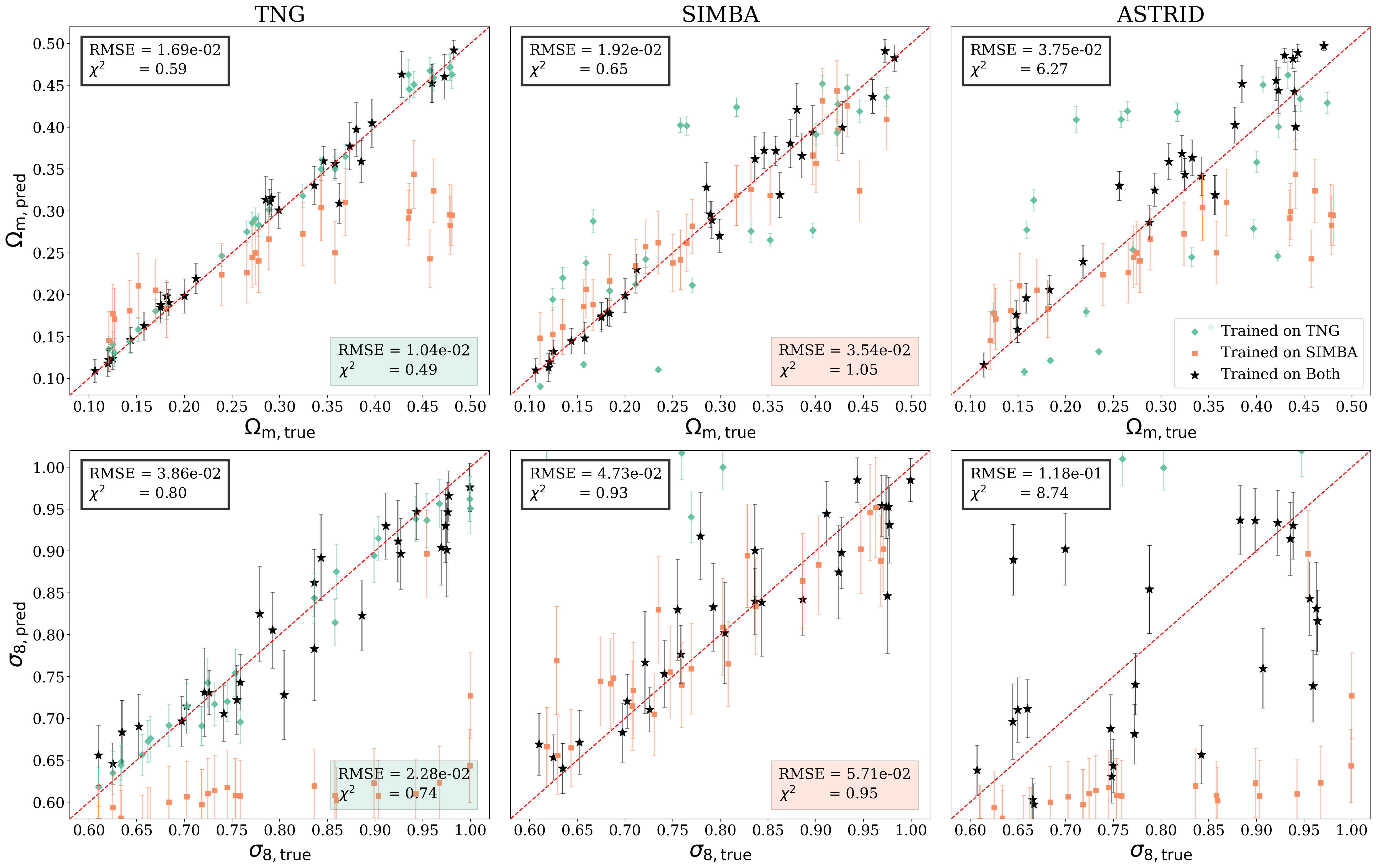}
    \caption{We train a neural network to infer $\Omega_m$ (upper panels) and $\sigma_8$ (lower panels) from gas temperature maps, tested on TNG (left column), SIMBA (middle column), and ASTRID (right column). The green, orange, and black colors represent the models trained on (the LH set of) TNG, SIMBA, and TNG+SIMBA suites respectively. The data points and error bars give the posterior mean and variance. The scores of RMSE and $\chi^2$ are shown in the legends.}
\label{fig:two-suite-test}
\end{figure*}

\begin{figure*}
\centering
\includegraphics[width=0.85\textwidth]{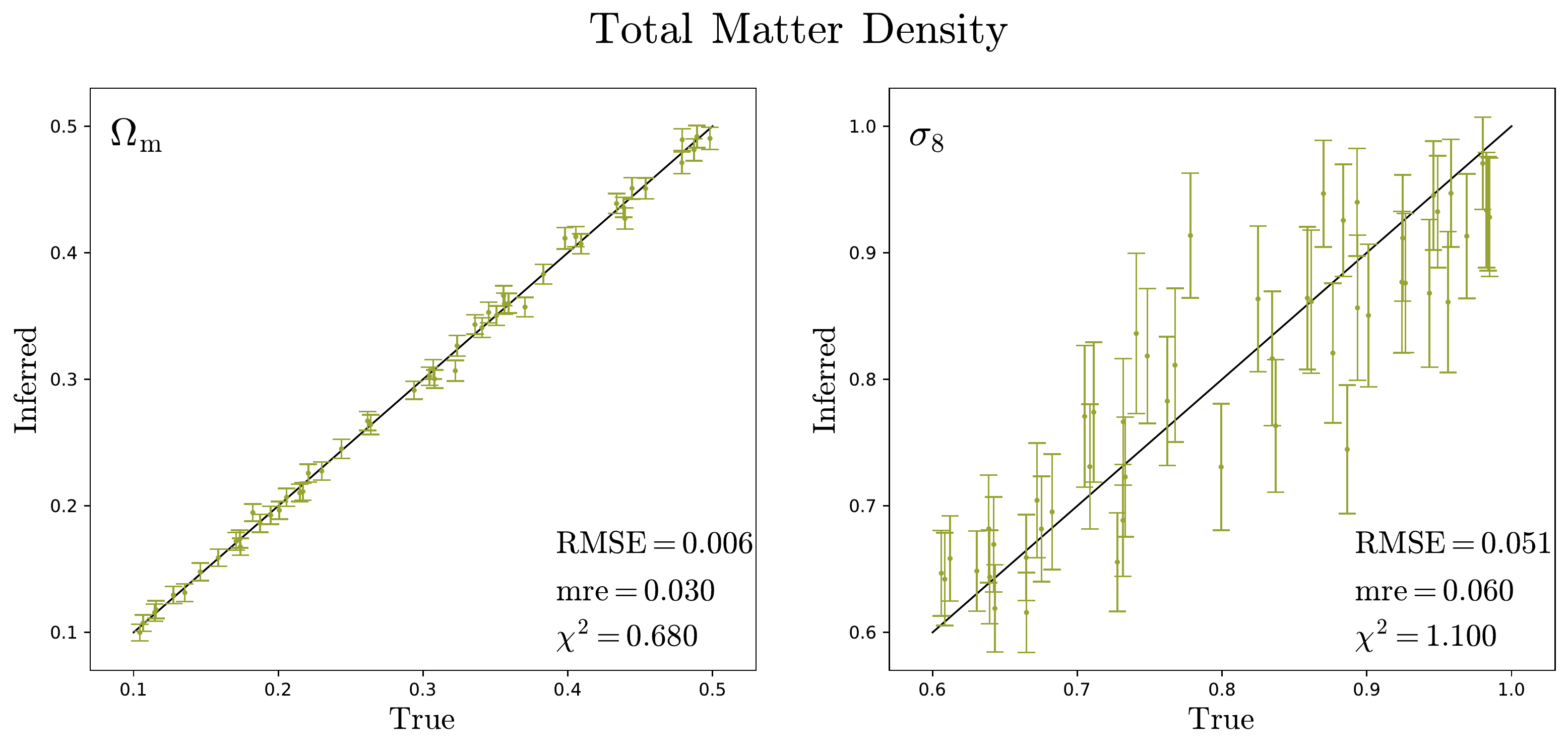}
\caption{We show the results of a neural network trained and tested on the TNG-SB28 set to infer $\Omega_m$ and $\sigma_8$ from 2D total matter density maps. This demonstrates that it is still possible to train an ML model to infer cosmological parameters even with the simulation set of TNG-SB28 that broadly (and sparsely) samples the high dimensional parameter space of a galaxy formation model.}
\label{fig:SB28_Mtot}
\end{figure*}

\subsection{Training on two simulation suites combined}
\label{subsec6.2:two-suite}

With a third suite of hydrodynamic simulations in CAMELS, we can train ML models on two simulation suites (i.e., based on two galaxy formation models) and test whether they can generalize to a third simulation suite that is carried out with a distinct galaxy formation model. 
We perform this task by training a CNN to infer $\Omega_m$ and $\sigma_8$ based on the 2D gas temperature map. 
The details of the ML model are presented in \cite{Paco2021-MultifieldCosmology}.
As shown in that paper, the ML model trained on TNG gas temperature maps does not generalize to SIMBA (and vice versa), due to the distinct astrophysical models utilized in each suite.

In this section, we train the ML model based on the combination of TNG and SIMBA maps and test that model on the ASTRID maps.
We note that the ML model is only trained and tested on the 2D gas temperature maps from the LH sets of the three simulation suites, to link to our previous work.
In Fig.~\ref{fig:two-suite-test}, the black points (with error bars) show the results of the ML model trained based on TNG+SIMBA maps and tested on TNG, SIMBA, and ASTRID individually. 
As a comparison, we also show the test results from the model trained only on TNG (green) and only on SIMBA (orange) that are presented in \cite{Paco2021-MultifieldCosmology}.
To qualify the model performance, we calculate the Root Mean Squared Error (RMSE) and reduced chi-squared ($\chi^2$) to quantify the precision of the inference model and the accuracy of the estimated errors. 

Based on Fig.~\ref{fig:two-suite-test}, and in particular the RMSE and $\chi^2$ scores, it is clear that while the models trained only on one suite perform badly on the other suite (exceedingly so for $\sigma_8$), the model trained on TNG+SIMBA performs well when tested on TNG or SIMBA. 
However, its performance degrades when tested on ASTRID, indicating that the model trained on two simulation suites does not generalize well when applied to a new galaxy formation model. 
Nevertheless, it is interesting to note that when tested on ASTRID, the model trained on TNG+SIMBA (black) outperforms the models trained only on TNG (green) or SIMBA (orange). 
One might worry that when training a model based on two distinct suites of gas temperature maps, it would categorize the input field as either TNG-like or SIMBA-like and make an inference that resembles the result from a model trained on one of the two separately. 
The fact that the black model outperforms both the orange and green models on ASTRID suggests that this is not the case. We also compared the inference results on individual ASTRID maps from the three ML models and found that in many cases, the black model can make more accurate inferences when the other two models are biased.

It is encouraging to see that the model trained on two suites is performing better than the ones trained on a single suite, and it shows some degree of extrapolation when applied to a new suite of galaxy formation models. 
Our speculation is that during the training process on the two distinct simulation suites, the ML model is encouraged to ignore features specific to TNG or SIMBA that make the single-suite-trained models less robust when tested on a previously-unseen suite. 
Instead, it focuses on capturing more general features that are shared between TNG and SIMBA to make the inference. 
Therefore, it seems promising to improve the robustness of ML models by including more distinct galaxy formation simulations in the training set. 

However, we should note that the black model's performance on ASTRID (especially on $\sigma_8$) is relatively poor compared to the results tested on TNG or SIMBA. 
Thus, we still caution that blindly training a model based on more simulation suites does not guarantee generalization to other galaxy formation models.
Specially-tailored machines that are trained with robustness in mind are a promising research direction, which we will explore in \cite{Jo-prep}.


\subsection{Training on the TNG-SB28 set}
\label{subsec6.3:TNG28-tests}

In this section, we present the first results of ML models trained on the new TNG-SB28 simulation set. The TNG-SB28 set sparsely samples a high-dimensional space of 28 cosmological and astrophysical parameters with 1024 simulations. 
It was a-priori uncertain whether it is feasible to train an ML model for cosmological inference based on such a simulation set or whether many more simulations would be needed to perform such task.

In Fig.~\ref{fig:SB28_Mtot}, we show the test results for TNG-SB28, where we train a neural network to infer $\Omega_m$ and $\sigma_8$ based on the 2D total mass density field generated from TNG-SB28. 
The model architecture is the same as that introduced in \cite{Paco_robust_maps}. 
The model yields good scores for $\Omega_m$ and worse, yet still constraining, results for $\sigma_8$, despite the broad variation (and possible degeneracy) of all the 28 parameters in TNG-SB28.

In our previous work \citep{Paco2021-MultifieldCosmology} training the neural network on the TNG-LH set and testing on the TNG-LH set, the mean relative error (mre) based on the 2D total matter density maps is $\langle \delta \Omega_m / \Omega_m \rangle = 0.034$ and $\langle \delta \sigma_8 / \sigma_8 \rangle = 0.024$.
Compared to the results based on the TNG-LH set, the constraining power on $\Omega_m$ does not degrade by making the parameter space larger in TNG-SB28 (with mre $\langle \delta \Omega_m / \Omega_m \rangle = 0.030$), while $\sigma_8$ is affected by it (the corresponding mre score degrades to $\langle \delta \sigma_8 / \sigma_8 \rangle = 0.060$), potentially due to the degeneracy introduced by additional parameters varied.
However, there is still some constraining power on $\sigma_8$. 
We have also conducted a similar training and testing for the $\Omega_m$ and $\sigma_8$ inference model based on gas temperature maps, which produced slightly worse but still constraining results, with mre $= 0.064$ and $0.072$ for $\Omega_m$ and $\sigma_8$ respectively (while the corresponding mre values based on the TNG-LH set are 0.053 and 0.037 for the gas temperature maps).

This section demonstrates the possibility of training ML models for cosmological inference based on the TNG-SB28 set. In future works, we will thoroughly investigate which categories of physical maps can infer which cosmological or astrophysical parameters in the TNG-SB28 set.

\section{Summary and Conclusion}
\label{sec:conclusions}

We have introduced the CAMELS-ASTRID simulation suite as the third large hydrodynamic simulation suite in the CAMELS project, carried out using the galaxy formation model of the ASTRID simulation. 
The core CAMELS-ASTRID suite contains the \{CV, 1P, LH, EX\} simulation sets with the same design as the TNG and SIMBA suites presented in \cite{Villaescusa-Navarro2021ApJ...915...71V}, exploring the cosmological and astrophysical parameter space of \{$\Omega_m$, $\sigma_8$, $A_{\rm SN1}$, $A_{\rm SN2}$, $A_{\rm AGN1}$, $A_{\rm AGN2}$ \}.
The extension of CAMELS-ASTRID contains an additional LHOb set also varying the cosmological parameter $\Omega_b$ along with the 6 fiducial parameters.

We have also presented CAMELS extension simulation sets based on TNG and SIMBA that aim to explore a higher dimensional parameter space in their respective galaxy formation models.  
In particular, the TNG-SB28 simulation set contains 1024 simulations that semi-uniformly sample a space of 5 cosmological parameters and 23 sub-grid parameters. 
It is accompanied by the TNG-1P-28 set that varies one parameter at a time to study their individual impact on different astrophysical properties. 
The analogous set SIMBA-1P-28 is also carried out to explore the high-dimensional parameter space in SIMBA.
Details of the 1P-28 sets in TNG and SIMBA are presented in the Appendix. 

We summarize the main features of the new simulation sets below.


\begin{itemize}

    \item The fiducial model of ASTRID exhibits the least impact on the large-scale gas properties and baryonic feedback on the matter power spectrum due to a milder AGN jet mode feedback compared to TNG and SIMBA.

    \item Varying feedback efficiency introduces non-linear effects to the overall galaxy and black hole populations and eventually the matter power spectrum.
    One example is the $A_{\rm AGN2}$ parameter in the ASTRID suite that controls the AGN thermal feedback efficiency. A larger $A_{\rm AGN2}$ suppresses the formation of massive black holes, curtails the energy budget of AGN feedback, brings less baryonic suppression on the total matter power spectrum, as well as enhances the global star formation rate and galaxy populations. 
    Given the intricacy of the convoluted effects of feedback processes, we should view the astrophysical parameters $A_{\rm SN}$ and $A_{\rm AGN}$ as modulations of various astrophysical processes that lead to variations in different physical quantities, rather than directly infer the amount of feedback from each astrophysical process as would be indicated by $A_{\rm SN}$ and $A_{\rm AGN}$ at face value.

    \item We show that in cases where the cosmological parameters are varied identically between the ASTRID, TNG, and SIMBA suites, the responses to those parameters can show strong differences between simulation models. 
    For example, the matter power spectrum responds differently to $\Omega_m$ and $\sigma_8$ on small scales between ASTRID, TNG, and SIMBA. 
    In the extended 1P-28 sets, TNG and SIMBA show different responses of gas power spectrum to changes in the $h$ and $n_s$ cosmological parameters.
    This stresses the importance of including different galaxy formation models to improve and test the robustness of cosmological inference tasks. 

    \item The TNG-1P-28 and SIMBA-1P-28 sets allow us to develop a more complete picture of the sensitivity of the galaxy formation process to a large number of parameter variations in the TNG and SIMBA models and their effects on the cosmic matter distribution. 
    We find that each of the 28 parameters, which are varied within reasonable ranges based on physical intuition, have significant effects on the results of the models. 
    Further, there is significant variability in the effect of different parameters on different quantities or `observables', where the overall impact of increasing one parameter is not necessarily linear or even monotonic. 
    This demonstrates the enormity of the overall relevant galaxy formation model parameter space, which is sampled much more extensively here than in previous work but it is still largely unexplored and not well-understood.

    \item Compared to the LH sets of TNG and SIMBA, the ASTRID-LH set drives larger variations in the gas power spectrum, the baryonic effect on the total matter power spectrum, the halo baryon fraction, the star formation rate density, and the galaxy population. 
    In the ASTRID suite, the matter power spectrum is sensitive to the $A_{\rm AGN2}$ and $A_{\rm SN2}$ parameters that regulate black hole growth as well as the AGN jet mode feedback.
    The large variation in the cosmic star formation rate and global galaxy properties is mainly driven by the sensitivity of ASTRID to the $A_{\rm SN2}$ parameter (which controls the SN wind speed).
    The broad variation of the ASTRID LH set in some aspects promotes the robustness of machine learning model training. 
    We find that machine learning models trained on galaxy catalogs from the ASTRID-LH set exhibit the best extrapolation performances when applied to other simulation sets.

    \item The new TNG-SB28 set spans a much larger parameter space of the TNG model compared to the original TNG-LH set, and indeed it produces wider variations in the simulation results in all the individual quantities and scaling relations that we have examined, typically variations that are roughly as wide as those from the ASTRID-LH set. The degree to which this extended TNG set produces more varied results than the original TNG-LH set is, however, different between different quantities. We hypothesize that in some cases this is due to an averaging effect of many important parameters, while in some cases the new set produces significantly larger variations than the original due to the introduction of new parameters that have a particularly significant effect of the quantity in question. The TNG-SB28 set probes a much more complete account of the flexibility in the TNG framework, and can be used as a more realistic testing ground for machine learning models, since it does not hold fixed the values of many parameters that are fundamentally not known but are kept fixed in the original TNG-LH set.

    
\end{itemize}

We have presented a few ML applications allowed by the new simulation sets introduced in this work, as summarized below:

\begin{itemize}
    \item 
    We evaluate the generalization performance of an ML inference model trained on 2D mass maps from the TNG suite, which generalizes to and can successfully predict $\Omega_m$ and $\sigma_8$ in the SIMBA suite. However, when tested on the ASTRID suite, the model fails to infer $\sigma_8$, despite the fact that ASTRID 2D mass maps are more similar to TNG in many statistical properties compared to SIMBA. This underscores the importance of including a diverse set of independent galaxy formation models to assess the robustness of trained ML models.

    \item 
    We develop a cosmological ML inference model by training on gas temperature maps, which have been previously shown to give rise to models that are not robust, from a combination of the TNG+SIMBA suites. 
    While when trained on TNG+SIMBA the resulting model generalizes well between TNG and SIMBA, when we test its ability to predict $\Omega_m$ and $\sigma_8$ on the ASTRID data, our findings suggest that training a model using multiple simulation suites does not necessarily result in generalization to other galaxy formation models. 
    Yet, we observed that the two-suite trained model (TNG+SIMBA) performed better when tested on the ASTRID data than the models trained only on one suite (TNG or SIMBA separately), indicating the potential for improving the robustness of ML models by incorporating additional distinct galaxy formation models in the training sets.

    \item 
    We demonstrate the feasibility of training neural networks to infer $\Omega_m$ and $\sigma_8$ using the TNG-SB28 simulation set, which sparsely samples the high-dimensional parameter space of a galaxy formation model. 
    Our results show that it is possible to train accurate ML models using this set, which can successfully infer cosmological parameters. 

\end{itemize}

\section*{Acknowledgements}

All the CAMELS simulations introduced in this work were run on the computing clusters at the Simons Foundation.
YN acknowledges support from McWilliams graduate fellowship and ITC postdoctoral fellowship. 
FV and SG acknowledge support from NSF grant AST-2108944. The Flatiron Institute is supported by the Simons Foundation. 
DAA acknowledges support by NSF grants AST-2009687 and AST-2108944, CXO grant TM2-23006X, Simons Foundation Award CCA-1018464, and Cottrell Scholar Award CS-CSA-2023-028 by the Research Corporation for Science Advancement. 
SB acknowledges funding supported by NASA-80NSSC22K1897. 
TDM and RACC acknowledge funding from the NSF AI Institute: Physics of the Future, NSF PHY-2020295, NASA ATP NNX17AK56G, and NASA ATP 80NSSC18K101.

\section*{Data Availability}
The code to reproduce the simulation is available at \url{https://github.com/MP-Gadget/MP-Gadget}, and continues to be developed.
Further technical details on the CAMELS simulations and instructions to download the data can be found in \url{https://camels.readthedocs.io} and \url{https://www.camel-simulations.org}.

\appendix

\section{Description of the 28 varied parameters}
\label{sec:appendix-28param}

Here we provide brief yet explicit and complete accounts of the 28 parameters that are varied in each of the new TNG and SIMBA sets discussed in this work. 
In addition to a concise description of the physical meaning of each parameter, we provide references to previous work that presented the parameter in the broader context of the model, as well as the range within which we vary the parameter value in this work. 
We also note whether the sampling of the parameter values within said range in this work is uniform in linear or logarithmic space.

\subsection{TNG 28-parameter variations}
\label{sec:appendix-2-tng}
\begin{enumerate}
    \item \texttt{Omega0} is the standard cosmological parameter $\Omega_m$, the $z=0$ cosmic matter density in units of the critical density. Here it is sampled linearly between $0.1$ and $0.5$, around the fiducial value of $0.3$.
    \item \texttt{Sigma8} is the standard cosmological parameter $\sigma_8$, the RMS of the $z=0$ linear overdensity in spheres of radius $8\hmpc$. Here it is sampled linearly between $0.6$ and $1$, around the fiducial value of $0.8$.
    \item \texttt{WindEnergyIn1e51erg} ($A_{\rm SN1}$) is a normalization factor for the energy in galactic winds per unit star-formation, and is denoted as $\bar{e}_w$ in Eq.~3 in \citet{PillepichA_18}. In the existing literature using CAMELS, including in this work, it is commonly referred to as $A_{\rm SN1}$, albeit with a ratio of $3.6$ between the two. Here it is sampled logarithmically between $0.9$ and $14.4$, around the fiducial value of $3.6$.
    \item \texttt{VariableWindVelFactor} ($A_{\rm SN2}$) is a normalization factor for the galactic wind speed, and is denoted as $\kappa_w$ in Eq.~1 in \citet{PillepichA_18}\footnote{Note that its description in Eq.~4 in \citet{Villaescusa-Navarro2021ApJ...915...71V} is incorrect in that $A_{\rm SN2}$ should have been located in front of $\kappa_w$ instead of the whole RHS.}. In the existing literature using CAMELS, including in this work, it is commonly referred to as $A_{\rm SN2}$, albeit with a ratio of $7.4$ between the two. Here it is sampled logarithmically between $3.7$ and $14.8$, around the fiducial value of $7.4$.
    \item \texttt{RadioFeedbackFactor} ($A_{\rm AGN1}$) is a normalization factor for the energy in AGN feedback, per unit accretion rate, in the low-accretion state, and is implemented as a pre-factor in front of the RHS of Eq.~8 in \citet{Weinberger2017}. In the existing literature using CAMELS, including in this work, it is commonly referred to as $A_{\rm AGN1}$. Here it is sampled logarithmically between $0.25$ and $4.0$, around the fiducial value of $1.0$.
    \item \texttt{RadioFeedbackReiorientationFactor} ($A_{\rm AGN2}$) is a normalization factor for the frequency of AGN feedback energy release events in the low-accretion state, and is denoted as $f_{\rm re}$ in Eq.~13 in \citet{Weinberger2017}. In the existing literature using CAMELS, including in this work, it is commonly referred to as $A_{\rm AGN2}$, albeit with a ratio of $20$ between the two. Here it is sampled logarithmically between $10$ and $40$, around the fiducial value of $20$.
    \item \texttt{OmegaBaryon} is the standard cosmological parameter $\Omega_b$, the $z=0$ cosmic baryon density in units of the critical density. Here it is sampled linearly between $0.029$ and $0.069$, around the fiducial value of $0.049$.
    \item \texttt{HubbleParam} is the standard Hubble constant, in units of $100\kms\mpc^{-1}$. Here it is sampled linearly between $0.4711$ and $0.8711$, around the fiducial value of $0.6711$.
    \item \texttt{n\_s} is the standard cosmological parameter $n_s$, the spectral index of the initial fluctuations\footnote{Note that since the power of the fluctuations at a scale of $8\hmpc$ serves as a pivot point that is set by \texttt{Sigma8}, most scales available in our $25\hmpc$ boxes are enhanced when \texttt{n\_s} is increased, with smaller scales enhanced more than larger ones.}. Here it is sampled linearly between $0.7624$ and $1.1624$, around the fiducial value of $0.9624$.
    \item \texttt{MaxSfrTimescale} is the timescale for star-formation at the density threshold of star-formation, and is denoted as $t^{\star}_0$ in Eq.~21 in \citet{SH03}. Here it is sampled logarithmically between $1.135\gyr$ and $4.54\gyr$, around the fiducial value of $2.27\gyr$.
    \item \texttt{FactorForSofterEQS} is an interpolation factor between the effective equation of state for star-forming gas obtained from the \citet{SH03} sub-grid model and an isothermal equation of state with $10^4K$, and is denoted as $q_{\rm EOS}$ in \citet{SpringelV_05}. Here it is sampled logarithmically between $0.1$ and $0.9$, around the fiducial value of $0.3$.
    \item \texttt{IMFslope} is the power-law index of the stellar initial mass function above $1\msun$ (where below that mass, it is kept fixed following \citet{ChabrierG_03a}). Here it is sampled linearly between $-2.8$ and $-1.8$, around the fiducial value of $-2.3$.
    \item \texttt{SNII\_MinMass\_Msun} is the lower threshold for the mass of a star that produces a supernova explosion, and is denoted $M_{\rm SNII,min}$ in \citet{Vogelsberger2013}. It affects both the available energy for galactic winds feedback (and in that sense is degenerate with \texttt{WindEnergyIn1e51erg}, or $A_{\rm SN1}$) and the effective (namely stellar-population-averaged) stellar mass return and enrichment. Here it is sampled linearly between $4\msun$ and $12\msun$, around the fiducial value of $8\msun$.
    \item \texttt{ThermalWindFraction} is the fraction of the galactic wind feedback energy that is injected thermally (the other component being kinetically), and is denoted as $\tau_w$ in \citet{PillepichA_18}. Here it is sampled logarithmically between $0.025$ and $0.4$, around the fiducial value of $0.1$.
    \item \texttt{VariableWindSpecMomentum} is a normalization factor for the specific momentum in galactic winds per unit star-formation, and is denoted as mom$_w$ in \citet{Vogelsberger2013}. Note that it affects the specific momentum of the winds only via their mass-loading factor and not via their speed. Here it is sampled linearly between $0$ and $4000\kms$. It is worth noting that the variations in this parameter differ from those of all the others in that the fiducial value of $0$ lies at the edge of the variation range rather than in its middle, for physical reasons (negative values are unphysical).
    \item \texttt{WindFreeTravelDensFac} sets the gas density around (collisionless) galactic wind particles at which they recouple back into the hydrodynamics. Here it is sampled logarithmically between $0.005$ and $0.5$, around the fiducial value of $0.05$, all in units of the density threshold for star-formation.
    \item \texttt{MinWindVel} is the minimum value imposed for the galactic wind speed, and is denoted as $v_{w,{\rm min}}$ in \citet{PillepichA_18}. Here it is sampled linearly between $150\kms$ and $550\kms$, around the fiducial value of $350\kms$.
    \item \texttt{WindEnergyReductionFactor} is a normalization factor for the energy of galactic winds at high metallicity compared to low metallicity, and is denoted $f_{w,Z}$ in \citet{PillepichA_18}. Here it is sampled logarithmically between $.0625$ and $1.0$, around the fiducial value of $0.25$.
    \item \texttt{WindEnergyReductionMetallicity} sets the metallicity at which the transition from high- to low-energy galactic winds occurs, and is denoted as $Z_{w,{\rm ref}}$ in \citet{PillepichA_18}. Here it is sampled logarithmically between $0.0005$ and $0.008$, around the fiducial value of $0.002$, in terms of fractional metal mass.
    \item \texttt{WindEnergyReductionExponent} controls the abruptness in metallicity of the transition between high- and low-energy galactic winds occurs, and is denoted as $\gamma_{w,Z}$ in \citet{PillepichA_18}. Here it is sampled linearly between $1$ and $3$, around the fiducial value of $2$.
    \item \texttt{WindDumpFactor} is the fraction of the metals in a star-forming cell getting ejected into a galactic wind that get deposited in neighboring star-forming cells prior to the ejection, and is denoted as $1-\gamma_w$ in \citet{PillepichA_18}. Here it is sampled linearly between $0.2$ and $1.0$, around the fiducial value of $0.6$.
    \item \texttt{SeedBlackHoleMass} is the mass of seed supermassive black holes, as described in \citet{Vogelsberger2013}. Here it is sampled logarithmically between $2.5\times10^5\msun$ and $2.5\times10^6\msun$, around the fiducial value of $8\times10^5\msun$.
    \item \texttt{BlackHoleAccretionFactor} is a normalization factor for the Bondi rate for the accretion onto supermassive black holes, and is implemented as a pre-factor in front of the RHS of Eq.~2 in \citet{Weinberger2017}. Here it is sampled logarithmically between $0.25$ and $4$, around the fiducial value of $1$.
    \item \texttt{BlackHoleEddingtonFactor} is a normalization factor for the limiting Eddington rate for the accretion onto supermassive black holes, and is implemented as a pre-factor in front of the RHS of Eq.~3 in \citet{Weinberger2017}. Here it is sampled logarithmically between $0.1$ and $10$, around the fiducial value of $1$.
    \item \texttt{BlackHoleFeedbackFactor} is a normalization factor for the energy in AGN feedback, per unit accretion rate, in the high-accretion state, and is denoted as $\epsilon_{\rm f,high}$ in Eq.~7 in  \citet{Weinberger2017}. Here it is sampled logarithmically between $0.025$ and $0.4$, around the fiducial value of $0.1$. 
    We note that this parameter is similar to the $A_{\rm AGN2}$ parameter in ASTRID suite.
    \item \texttt{BlackHoleRadiativeEfficiency} is the radiative efficiency of AGN feedback, namely the fraction of the accretion rest-mass that is released in the accretion process. In the high-accretion state, it is denoted as $\epsilon_{\rm r}$ in Eq.~7 in \citet{Weinberger2017}. In the low-accretion state, it plays a dual role in setting $\epsilon_{\rm f,kin}$ in Eq.~9 in \citet{Weinberger2017} by: {\it i)} serving as an upper limit on its the value (replacing the constant 0.2 that appears in that equation), and {\it ii)} modulating the density at which that upper limit is reached, by scaling the density in the numerator of the first term in that equation, where a factor (\texttt{BlackHoleRadiativeEfficiency}/0.2) is missing for the general case of \texttt{BlackHoleRadiativeEfficiency}$\neq0.2$.
    Here it is sampled logarithmically between $0.05$ and $0.8$, around the fiducial value of $0.2$.
    \item \texttt{QuasarThreshold} is the Eddington ratio (at the `pivot mass' of $10^8\msun$) that serves as the threshold between the low-accretion and high-accretion states of AGN feedback, and is denoted as $\chi_0$ in Eq.~5 in \citet{Weinberger2017}. Here it is sampled logarithmically between $.000063$ and $.063$, around the fiducial value of $0.002$.
    \item \texttt{QuasarThresholdPower} is the power-law index of the scaling of the low- to high-accretion state threshold with black hole mass, and is denoted as $\beta$ in Eq.~5 in \citet{Weinberger2017}. Here it is sampled linearly between $0$ and $4$, around the fiducial value of $2$.
\end{enumerate}

\subsection{SIMBA 28-parameter variations}
\label{sec:appendix-2-simba}
\begin{enumerate}

\item \texttt{Omega0} is the $z=0$ cosmic matter density in units of the critical density. Here it is sampled linearly between $0.1$ and $0.5$ around the fiducial value $\Omega_{\rm m} =0.3$.
  
\item \texttt{Sigma8} is the RMS of the $z=0$ linear overdensity in spheres of radius $8\hmpc$. Here it is sampled linearly between $0.6$ and $1$ around the fiducial value of $\sigma_8 = 0.8$.

\item \texttt{$A_{\rm SN1}$} is a normalization factor for the mass loading of galactic winds relative to the scaling from FIRE simulations \citep{Angles-Alcazar2017_BaryonCycle} implemented in SIMBA. This is one of the feedback parameters originally varied in CAMELS \citep[see Eq.~7 in ][]{Villaescusa-Navarro2021ApJ...915...71V} and it is sampled logarithmically between $0.25$ and $4.0$ around the fiducial value $A_{\rm SN1} = 1.0$.

\item \texttt{$A_{\rm SN2}$} is a normalization factor for the speed of galactic winds relative to the scaling from FIRE simulations \citep{Muratov2015} implemented in SIMBA. This is one of the feedback parameters originally varied in CAMELS \citep[see Eq.~8 in ][]{Villaescusa-Navarro2021ApJ...915...71V} and it is sampled logarithmically between $0.5$ and $2.0$ around the fiducial value $A_{\rm SN2} = 1.0$.

\item \texttt{$A_{\rm AGN1}$} is a normalization factor for the momentum flux of kinetic AGN-driven outflows \citep{Angles-Alcazar2017_BHfeedback} in the radiative and jet modes. This is one of the feedback parameters originally varied in CAMELS \citep[see Eq.~9 in ][]{Villaescusa-Navarro2021ApJ...915...71V} and it is sampled logarithmically between $0.25$ and $4.0$ around the fiducial value $A_{\rm AGN1} = 1.0$.

\item \texttt{$A_{\rm AGN2}$} is a normalization factor for the speed of AGN-driven outflows in jet mode. This is one of the feedback parameters originally varied in CAMELS \citep[see Eq.~10 in ][]{Villaescusa-Navarro2021ApJ...915...71V} and it is sampled logarithmically between $0.5$ and $2.0$ around the fiducial value $A_{\rm AGN2} = 1.0$.

\item \texttt{OmegaBaryon} is the $z=0$ cosmic baryon density in units of the critical density. Here it is sampled linearly between $0.029$ and $0.069$ around the fiducial value of $\Omega_{\rm b} = 0.049$.
   
\item \texttt{HubbleParam} is the standard Hubble constant in units of $100\kms\mpc^{-1}$. Here it is sampled linearly between $0.4711$ and $0.8711$ around the fiducial value $h = 0.6711$.
   
\item \texttt{n\_s} is the spectral index of the initial fluctuations. Here it is sampled linearly between $0.7624$ and $1.1624$ around the fiducial value of $n_{\rm s} = 0.9624$.

\item \texttt{SfrCritDens} is the gas number density threshold for star formation. Here it is sampled logarithmically between 0.02\,cm$^{-3}$ and 2\,cm$^{-3}$ around the fiducial value of 0.2\,cm$^{-3}$.

\item \texttt{SfrEfficiency} is the star formation efficiency per free fall time, denoted as $\epsilon_\star$ in \citet{Dave2019MNRAS.486.2827D}.  Here it is sampled logarithmically between 0.01 and 0.04 around the fiducial value $\epsilon_\star = 0.02$. 

\item \texttt{ISMJeansFac} determines the level of artificial pressurization in the ISM such that the Jeans mass is resolved with a number of gas resolution elements given by \texttt{ISMJeansFac} $\times N_{\rm ngb}$, where $N_{\rm ngb} = 64$ is the number of neighbors in the smoothing kernel \citep{Dave2016}. Here \texttt{ISMJeansFac} is sampled logarithmically between $0.25$ and $4.0$ around the fiducial value of 1.0.

\item \texttt{WindTravTime} is the maximum amount of time that galactic winds can propagate before recoupling hydrodynamically to the surrounding gas, in units of the Hubble time at launch \citep{Dave2016}. Here it is sampled logarithmically between 0.2 and 0.002 around the fiducial value of $0.02$.

\item \texttt{WindDensFac} defines a gas number density threshold below which decoupled galactic winds are forced to recouple hydrodynamically to the surrounding gas \citep{Dave2016}.  Here it is sampled logarithmically between 0.1\,cm$^{-3}$ and 0.001\,cm$^{-3}$ around the fiducial value of 0.01\,cm$^{-3}$.

\item \texttt{WindColdTemp} defines the temperature assumed for the cold phase of galactic winds \citep{Dave2016}. Here it is sampled logarithmically between 100\,K and $10^{4}$\,K around the fiducial value of 1000\,K.

\item \texttt{WindHotFrac} is the fraction of wind particles ejected in the hot phase, where the wind temperature is determined by the fraction of the SN energy not used for kinetic ejection \citep{Dave2016}. Here \texttt{WindHotFrac} is sampled linearly between 0.0 and 0.6 around the fiducial value of 0.3.

\item \texttt{WindVelSlope} sets the power-law dependence of the galactic wind velocity on the circular veloscity of the galaxy such that ${\rm log_{10}}(v_{\rm wind}) \propto (1+\texttt{WindVelSlope}) \times {\rm log_{10}}(v_{\rm circ}$), as defined in Eq.~3 of
\citet{Dave2019MNRAS.486.2827D}. Here \texttt{WindVelSlope} is sampled linearly between -0.38 and 0.62 around the fiducial value of 0.12.

\item \texttt{AGBWindHeatVel} is the velocity of winds from AGB stars which are assumed to thermalize with the ambient ISM following the model of \citet{Conroy2015}. Here \texttt{AGBWindHeatVel} is sampled logarithmically between 25\,km\,s$^{-1}$ and 400\,km\,s$^{-1}$ around the fiducial value of 100\,km\,s$^{-1}$.

\item \texttt{BHSeedMass} is the mass of the supermassive black hole seeds, which is varied logarithmically between $10^{3}\,\hmsun$ and $10^{5}\,\hmsun$ around the fiducial value of $10^{4}\,\hmsun$ \citep{Dave2019MNRAS.486.2827D}.

\item \texttt{BHSeedRatio} defines the minimum galaxy stellar mass required for black hole seeding as $\texttt{BHSeedRatio} \times M_{\rm seed}$ \citep{Angles-Alcazar2017_BHfeedback}.  Here \texttt{BHSeedRatio} is sampled logarithmically between $3\times10^4$ and $3\times10^6$ around the fiducial value of $3\times10^5$.

\item \texttt{BHAccrFac} is a normalization factor for the black hole growth rate applied to both accretion of hot gas following \citet{Bondi1952} and accretion of cold gas driven by gravitational torques \citep{Hopkins2011_Analytic,Angles-Alcazar2017_BHfeedback}. Here \texttt{BHAccrFac} is sampled logarithmically between 0.25 and 4.0 around the fiducial value of 1.0.

\item \texttt{BHAccrMaxR} is the maximum size of the black hole kernel used when searching for neighboring gas elements. Here it is sampled logarithmically between 2\,$\hkpc$ and 8\,$\hkpc$ around the fiducial value of 4\,$\hkpc$.

\item \texttt{BHAccrTempThr} is the temperature threshold above which gas accretes onto the black hole at the Bondi rate and below which gas accretion proceeds at the gravitational torque accretion rate \citep{Dave2019MNRAS.486.2827D}. Here it is sampled logarithmically between $10^4$\,K and $10^6$\,K around the fiducial value of $10^5$\,K.

\item \texttt{BHEddingtonFac} is a normalization factor for the limiting Eddington rate applied to the accretion of cold gas onto black holes through gravitational torques (hot-mode accretion is always limited to the Eddington rate). Here \texttt{BHEddingtonFac} is sampled logarithmically between 0.75 and 12.0 around the fiducial value of 3.0.

\item \texttt{BHNgbFac} sets the desired effective number of gas resolution elements within the black hole kernel as \texttt{BHNgbFac} $\times N_{\rm ngb}$, where $N_{\rm ngb} = 64$ is the number of neighbors in the gas smoothing kernel. Here it is sampled linearly from 2 to 6 around the fiducial value of 4.

\item \texttt{BHRadiativeEff} is the radiative efficiency of the black hole accretion disk, denoted as $\eta$ in \citet{Dave2019MNRAS.486.2827D}, which determines the fraction of the rest-mass energy of the accreting material lost to radiation, the amount of energy injected as X-ray feedback, and the momentum flux of outflows in radiative and jet modes \citep{Dave2019MNRAS.486.2827D}. Here \texttt{BHRadiativeEff} is sampled logarithmically from 0.025 to 0.4 around the fiducial value of 0.1.

\item \texttt{BHJetTvirVel} sets the outflow velocity above which gas ejected in jet mode is heated to the virial temperature of the host halo \citep{Dave2019MNRAS.486.2827D}. Here \texttt{BHJetTvirVel} is sampled logarithmically from 500\,km\,s$^{-1}$ to 8000\,km\,s$^{-1}$ around the fiducial value of 2000\,km\,s$^{-1}$.

\item \texttt{BHJetMassThr} is the black hole mass threshold above which black holes are allowed to transition into the jet feedback mode. It is sampled logarithmically from $4.5\times10^6\,M_\odot$ to $4.5\times10^8\,M_\odot$ around the fiducial value of $4.5\times10^7\,M_\odot$ \citep{Dave2019MNRAS.486.2827D}.

\end{enumerate}

\section{TNG extended 1P set}
\label{sec:appendix-TNG-1P}

\begin{figure*}
\centering
\includegraphics[width=1.0\textwidth]{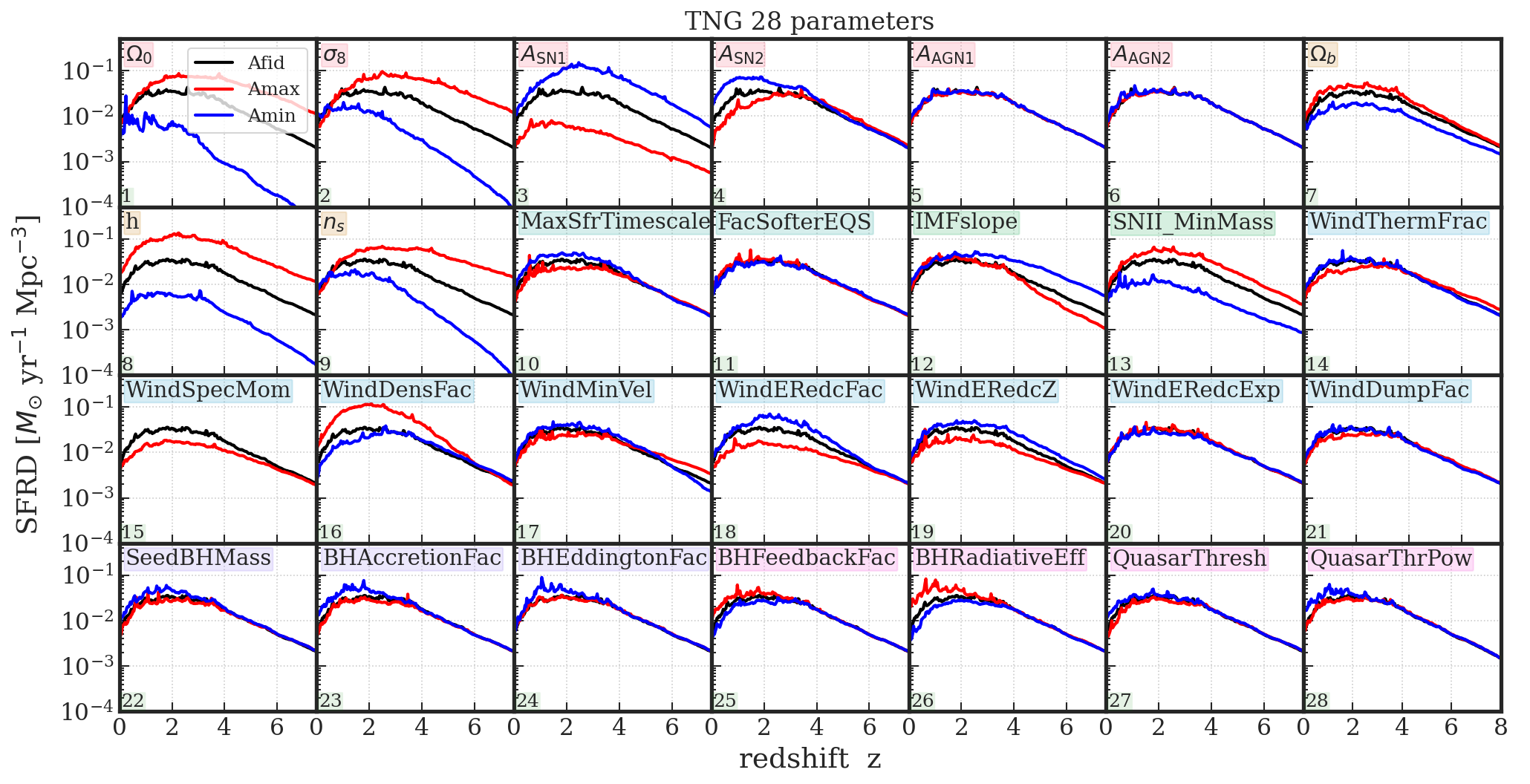}
\caption{Global star formation rate density of the TNG extended 1P set. The 28 panels reproduce the fiducial model as the black line. The red and blue lines in each panel represent the simulations run with the highest and lowest parameter value (except for the \texttt{WindSpecMom} parameter in TNG, see text for more details).
}
\label{fig:SB1P_SFRD}
\end{figure*}

\begin{figure*}
\centering
\includegraphics[width=1.0\textwidth]{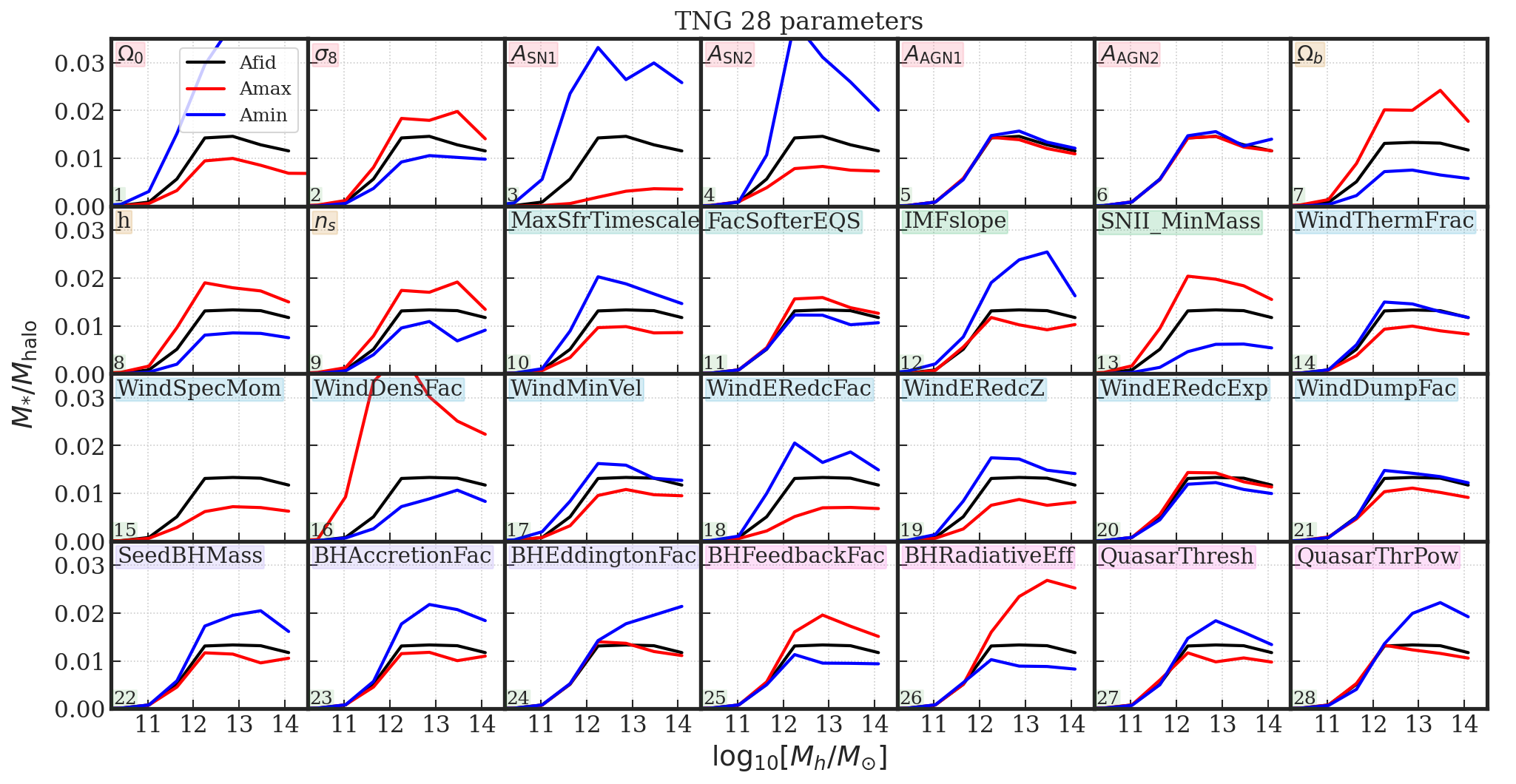}
\caption{Same as Fig.~\ref{fig:SB1P_SFRD}, this plot gives the $z=0$ stellar mass fraction of the TNG extended 1P set. $M_{\rm halo}$ here is the FOF halo mass, and $M_*$ is the stellar mass within the FOF halo. $y$ axis gives the averaged ($M_*/M_{\rm halo}$) for each halo mass bin for each simulation.}
\label{fig:SB1P_sfrac}
\end{figure*}

\begin{figure*}
\centering
\includegraphics[width=1.0\textwidth]{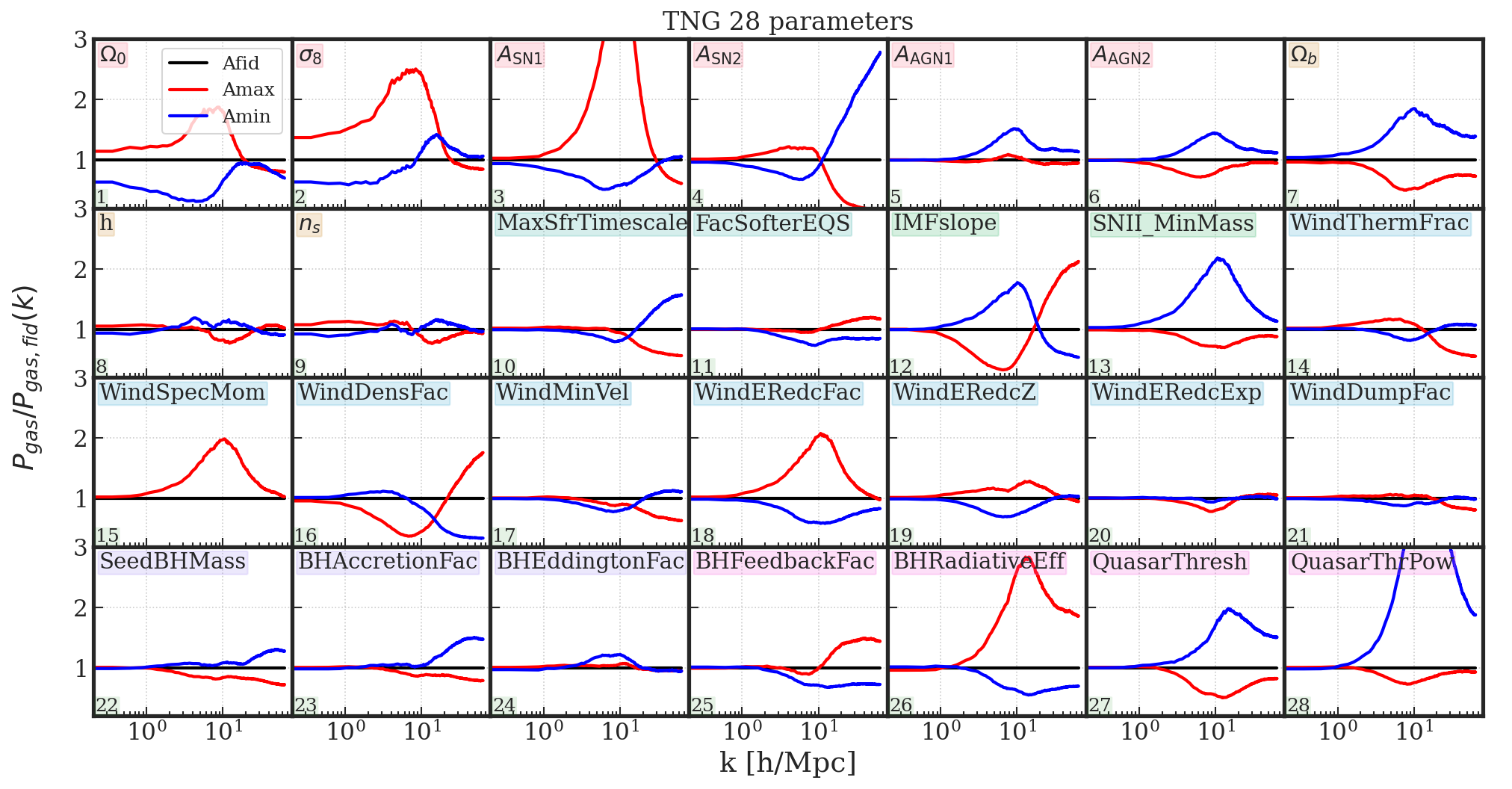}
\caption{Same as Fig.~\ref{fig:SB1P_SFRD}, this plot gives the ratio between the gas power spectrum of the TNG extended 1P simulations and that of the fiducial model.}
\label{fig:SB1P_gaspow}
\end{figure*}

Here we present a few examples for how individual model parameters affect the results of simulations in the TNG framework, by way of showing three distinct quantities based on the TNG-1P-28 set -- the cosmic SFRD, the $M_{\rm BH} - M_{*}$ relation and the gas power spectrum $P_{\rm gas}(k)$ normalized by that of the fiducial model. Figs.~\ref{fig:SB1P_SFRD}, \ref{fig:SB1P_sfrac} and \ref{fig:SB1P_gaspow} each shows one of these quantities, respectively, for 57 simulations from the TNG-1P-28 set, which are the two extremes for each of the 28 parameters (red for the highest and blue for the lowest value) as well as the fiducial model (black, repeating in all panels). The panels are organized such that the first 6 correspond to the original 6 parameters from the TNG-LH set, following by the 3 new cosmological parameters, followed by the star-formation and then the galactic winds parameters, and finally in the bottom row the black hole growth and AGN feedback parameters.

This presentation of the dependencies of these three quantities on the model parameters in Figs.~\ref{fig:SB1P_SFRD}, \ref{fig:SB1P_sfrac} and \ref{fig:SB1P_gaspow} serves both to support the interpretation of the diversity of results in the TNG-SB28 set that is discussed in Section \ref{section: CVLH}, to contribute to the justification for the choice of parameter variation ranges, and to explore the degree to which the various parameters are degenerate with one another in terms of their effects on the simulation results, as discussed below.

Fig.~\ref{fig:SB1P_SFRD}, for the cosmic SFRD versus redshift, demonstrates that some of the original 6 parameters, namely $\Omega_m$, $\sigma_8$ and $A_{\rm SN1}$ are some of the most influential ones. However, two of the new cosmological parameters, $h$ and $n_s$, have effects of similar overall scale. It is notable, though, that the response to each of these 5 most influential parameters has a unique shape, such that they are not degenerate with one another, even for just this single quantity. A few additional galactic winds parameters also have fairly significant effects; it is worth noting that one of those, \texttt{WindFreeTravelDensFac}, can be considered rather `numerical' in nature as it controls the ambient density at which collisionless `wind particles' recouple into the hydrodynamics. This highlights a situation where even `nuisance' parameters without a very clear analog in physical reality, which are often `buried' deep inside the model, can be consequential for modern galaxy formation simulations. Finally, we note that black hole parameters have a mild effect on this quantity, as expected, given that they tend to affect the most massive galaxies but that star-formation is dominated by low- and medium-mass galaxies.

Fig.~\ref{fig:SB1P_sfrac}, for the $z=0$ stellar-to-halo mass ratio, shows a much broader set of parameters that matter than is the case for the SFRD in Fig.~\ref{fig:SB1P_SFRD}. Essentially all the parameters except the original AGN parameters $A_{\rm AGN1}$ and $A_{\rm AGN2}$ make a visible effect in the figure. On the other hand, of the galactic winds parameters, the original ones, $A_{\rm SN1}$ and $A_{\rm SN2}$, are the most significant, joined by \texttt{WindFreeTravelDensFac}. These types of parameters, as well as the cosmological ones, tend to affect a wide range of halo masses, while the black hole and AGN parameters meaningfully affect only halos more massive than $\sim10^{12}\msun$, as expected. Newly varied star-formation parameters, the star-formation timescale and the IMF slope, also have a significant impact. 
It is interesting to explore the degeneracy, and breaking thereof, across the two quantities shown in Figs.~\ref{fig:SB1P_SFRD} and \ref{fig:SB1P_sfrac}. For example, the effects of $n_s$ and \texttt{MaxSfrTimescale} (second and third panels on the second row) appear to be almost degenerate for the $M_*$-$M_h$ relation, but they have radically different consequences for the SFRDs. 
Conversely, $A_{\rm SN2}$ and \texttt{BlackHoleRadiativeEfficiency}, which have a similar effect on the SFRDs, have rather distinct impacts on the $M_*$-$M_h$ relation.

Finally, Fig.~\ref{fig:SB1P_gaspow}, for the gas power spectrum $P_{\rm gas}(k)$ (relative to the fiducial model), demonstrates its own distinctive trends. For example, the newly added cosmological parameters $h$ and $n_s$ that prove so significant for the SFRDs in Fig.~\ref{fig:SB1P_SFRD}, make barely a dent on $P_{\rm gas}(k)$. In contrast, \texttt{QuasarThresholdPower}, which does not stand out as the most important AGN parameter in Figs.~\ref{fig:SB1P_SFRD} and \ref{fig:SB1P_sfrac}, has a very strong effect on $P_{\rm gas}(k)$, in particular when its value is small, likely by moving many massive black holes from the kinetic to the thermal feedback mode. Overall, $P_{\rm gas}(k)$ shows a large variation in terms of the strength of its dependence on different parameters, while both the important ones and the non-important ones each sample all parameter types: cosmological, star-formation and ISM, galactic winds, and black hole parameters.

\section{SIMBA extended 1P set}
\label{sec:appendix-SIMBA-1P}

\begin{figure*}
\centering
\includegraphics[width=1.0\textwidth]{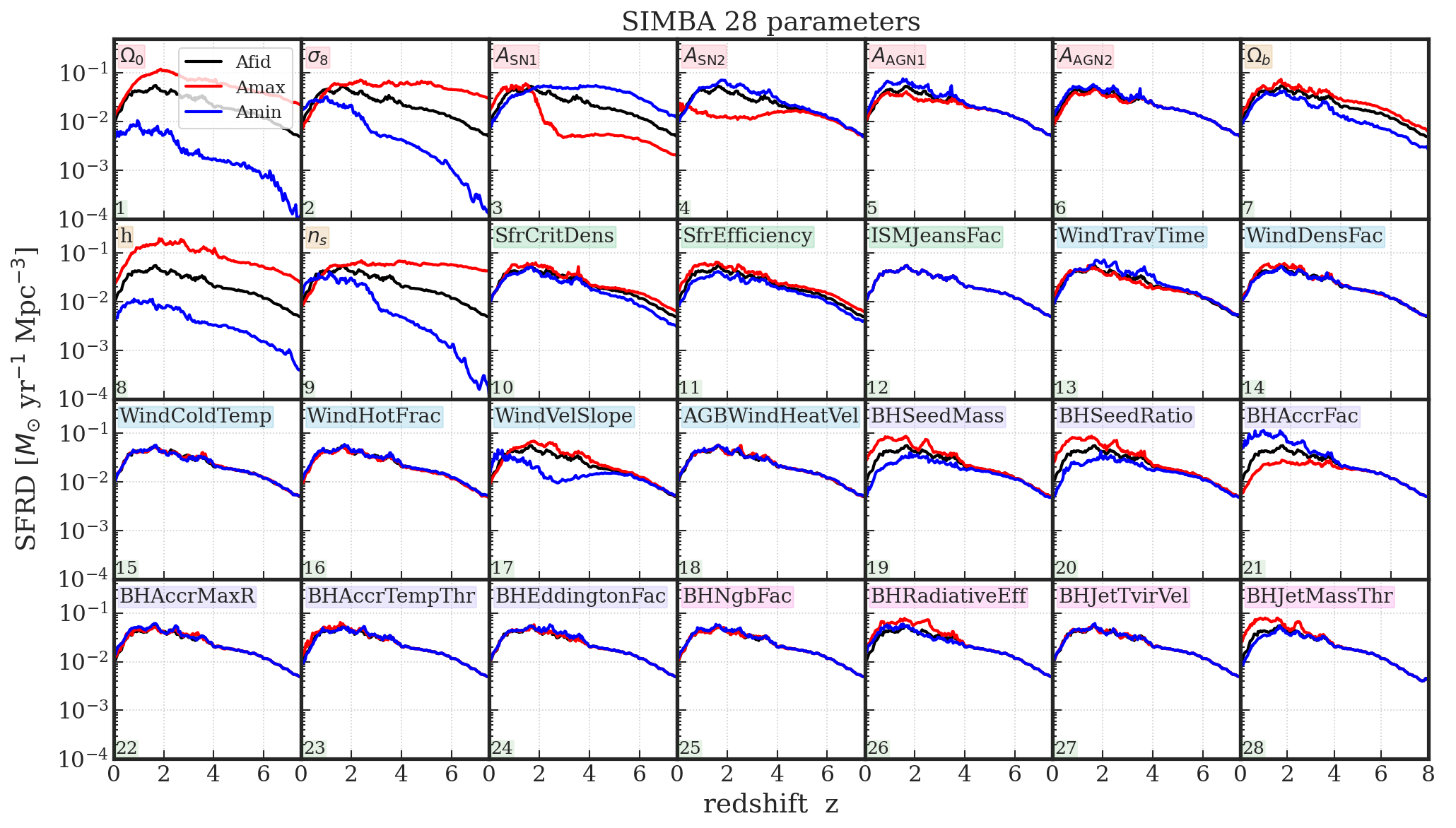}
\caption{Global star formation rate density of the SIMBA extended 1P set. The 28 panel shares the same black line that corresponds to the result from the fiducial model of SIMBA. }
\label{fig:SIMBA_SB1P_SFRD}
\end{figure*}

\begin{figure*}
\centering
\includegraphics[width=1.0\textwidth]{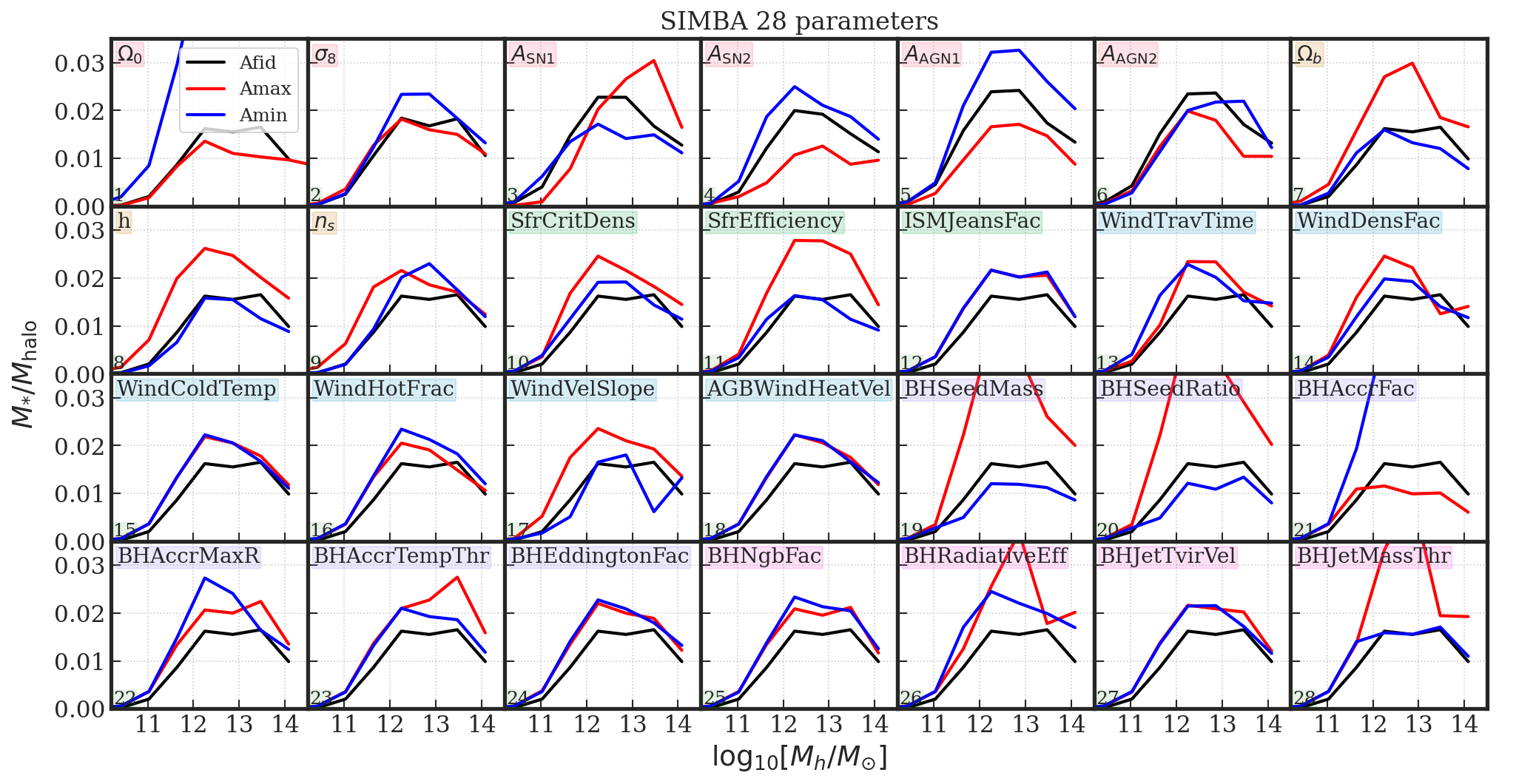}
\caption{Same as Fig.~\ref{fig:SIMBA_SB1P_SFRD}, this plot gives the $z=0$ stellar mass fraction of the SIMBA extended 1P set. $M_{\rm halo}$ here is the FOF halo mass, and $M_*$ is the stellar mass within the FOF halo. $y$ axis gives the averaged ($M_*/M_{\rm halo}$) for each halo mass bin for each simulation.}
\label{fig:SIMBA_SB1P_sfrac}
\end{figure*}

\begin{figure*}
\centering
\includegraphics[width=1.0\textwidth]{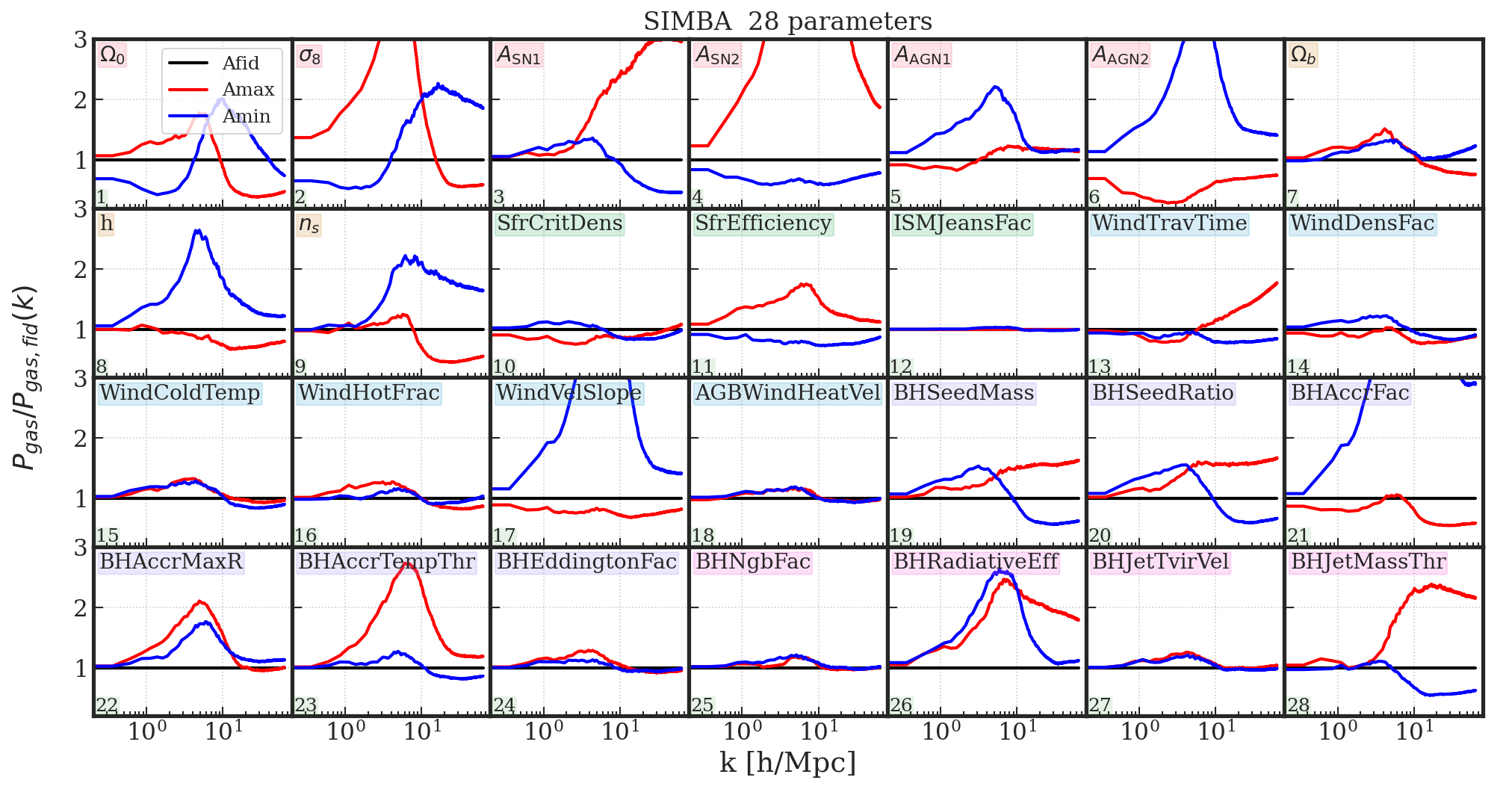}
\caption{Same as Fig.~\ref{fig:SIMBA_SB1P_SFRD}, this plot gives the ratio between the gas power spectrum of the SIMBA extended 1P simulations and that of the fiducial model.}
\label{fig:SIMBA_SB1P_gaspow}
\end{figure*}

As in Appendix~\ref{sec:appendix-TNG-1P} for TNG, here we illustrate how individual parameters in the SIMBA model affect global properties of galaxies and the large scale distribution of baryons in simulations. 
Fig.~\ref{fig:SIMBA_SB1P_SFRD} shows the evolution of the cosmic SFRD, Fig.~\ref{fig:SIMBA_SB1P_sfrac} shows the stellar mass--halo mass relation at $z=0$, and  
Fig.~\ref{fig:SIMBA_SB1P_gaspow} shows the gas power spectrum $P_{\rm gas}(k)$ at $z=0$ normalized by that of the fiducial model. Each figure consists of 28 panels corresponding to all parameter variations in the SIMBA-1P-28 set, where the fiducial model is indicated by the black line, and the red and blue lines correspond to the results for the highest and lowest parameter values, respectively. The first six panels correspond to the original SIMBA parameter variations in CAMELS \citep{Villaescusa-Navarro2021ApJ...915...71V}, including $\Omega_{\rm m}$, $\sigma_8$, and four parameters controlling the efficiency of kinetic outflows driven by stellar feedback ($A_{\rm SN1}$ and $A_{\rm SN2}$) and AGN feedback ($A_{\rm AGN1}$ and $A_{\rm AGN2}$).  Then we show variations of three additional cosmological parameters ($\Omega_{\rm b}$, $h$, and $n_{\rm s}$) followed by astrophysical parameters controlling star formation, galactic winds, black hole accretion, and AGN feedback.

Fig.~\ref{fig:SIMBA_SB1P_SFRD} shows that the cosmic SFRD increases systematically with higher $\Omega_{\rm b}$, $h$, and $n_{\rm s}$, as seen for TNG in Fig.~\ref{fig:SB1P_SFRD}.  
Interestingly, the SFRD increases not only with the star formation efficiency, as expected, but also with a higher threshold density for star formation, suggesting that stellar feedback is less efficient at regulating star formation when the star-forming gas reaches higher densities.  Besides the normalization of the mass loading ($A_{\rm SN1}$) and velocity ($A_{\rm SN2}$) of galactic winds, the next most important wind parameter appears to be the power-law dependence of wind speed with the circular velocity of the galaxy \texttt{WindVelSlope}. Numerical parameters related to the hydrodynamic treatment of winds also have a noticeable effect, with longer hydrodynamic decoupling times decreasing the SFRD and higher recoupling densities having the opposite effect.
Interestingly, increasing the BH seed mass at fixed BH-to-host mass ratio has similarly strong effects on the SFRD as increasing the BH-to-host mass ratio by the same factor at fixed BH seed mass.  This suggests that the seed mass itself is not important, as expected given the weak dependence of gravitational torque accretion on BH mass \citep{Angles-Alcazar2013,Angles-Alcazar2017_BHfeedback}, but the stellar mass at which galaxies are seeded is crucial, with significantly higher SFRD when BH seeding occurs only in higher mass galaxies.  Relatedly, increasing the BH accretion normalization systematically decreases the SFRD.
Other important AGN feedback parameters besides the momentum flux ($A_{\rm AGN1}$) and the jet speed ($A_{\rm AGN2}$) include the BH mass threshold for jet feedback, with higher SFRD when only the most massive BHs are allowed to transition to the jet mode.

Fig.~\ref{fig:SIMBA_SB1P_sfrac} shows that most parameter variations have a noticeable effect in the stellar mass--halo mass relation at $z=0$, as seen for TNG.  Increasing $\Omega_{\rm m}$ at fixed $\Omega_{\rm b}$ decreases the stellar-to-halo mass ratio, as expected given the decline in the cosmic baryon fraction \citep{Delgado2023}, while increasing $\Omega_{\rm b}$ at fixed $\Omega_{\rm m}$ has the opposite effect. Increasing the momentum flux of AGN-driven outflows ($A_{\rm AGN1}$) results in a reduction of $M_{*}/M_{\rm halo}$ across the halo mass range, while increasing the jet speed ($A_{\rm AGN2}$) only affects the higher mass galaxies \citep{Dave2019MNRAS.486.2827D}.
Interestingly, increasing the speed of galactic winds ($A_{\rm SN2}$) drives a clear decrease in $M_{*}/M_{\rm halo}$ at all masses, while increasing the mass loading of winds ($A_{\rm SN1}$) increases the stellar-to-halo mass ratio in massive galaxies. 
This counterintuitive effect could result from an increase in the intergalactic transfer of gas from lower mass galaxies, providing an additional source of gas to higher mass galaxies \citep{Angles-Alcazar2017_BaryonCycle}, or could reflect the complex non-linear interplay between stellar and AGN feedback as noted in previous works (\citealt{BoothSchaye2013,vanDaalen2020,Nicola2022,Delgado2023}; Gebhardt et al., in prep.).
As expected from Fig.~\ref{fig:SIMBA_SB1P_SFRD}, increasing the galaxy stellar mass above which BHs are seeded (i.e. higher \texttt{BHSeedMass} or \texttt{BHSeedRatio}) results in significantly higher $M_{*}/M_{\rm halo}$ across the halo mass range, while increasing the BH accretion normalization decreases $M_{*}/M_{\rm halo}$. Another expected trend is the strong increase in stellar mass at fixed halo mass when only the most massive black holes are allowed to switch into the jet feedback mode \citep{Dave2019MNRAS.486.2827D}.  Nonetheless, several parameter variations indicate complex non-linear behavior, with non-monotonic trends when systematically increasing parameter values.   

Fig.~\ref{fig:SIMBA_SB1P_gaspow} illustrates the very complex response of the gas power spectrum in SIMBA to changes in model parameters, where the relative impact of either increasing or decreasing parameter values often depends on and can even revert trend with scale. In addition to the original six parameters (1--6) varied in CAMELS, other key parameters affecting the clustering of gas include \texttt{WindVelSlope}, \texttt{BHSeedRatio}, \texttt{BHAccFac}, and \texttt{BHJetMassThr}, which also have important effects on the cosmic SFRD and the $M_{*}$--$M_{\rm halo}$ relation. Interestingly, increasing the threshold temperature for accretion of ``hot'' gas via Bondi accretion (instead of ``cold'' gas via gravitational torques) had only a minor impact on the SFRD but increased the $M_{*}/M_{\rm halo}$ ratio and therefore decreased the efficiency of AGN feedback in the most massive halos, resulting in enhanced clustering of gas on scales $k \lesssim 1\,h$\,Mpc$^{-1}$ relative to the fiducial model. 
It is also interesting to note that the variations of $h$ and $n_{\rm s}$ in SIMBA have a significant impact on the gas power spectrum while for TNG their effect on $P_{\rm gas}(k)$ was negligible, despite $h$ and $n_{\rm s}$ having similar large effects on the cosmic SFRD for both models. 
In SIMBA, the impact on $P_{\rm gas}(k)$ is primarily driven by AGN jet feedback, which comes from the most massive BHs whose growth and abundance is very sensitive to cosmology.
The fact that different simulation models predict different responses of the gas (and hence the total matter) power spectrum to the same variation of the cosmological parameters stresses the importance of incorporating diverse galaxy formation models to improve and test the robustness of cosmological inference.

\bibliography{references}{}
\bibliographystyle{aasjournal}

\end{document}